\documentclass[useAMS,usenatbib]{mn2e}
\usepackage{natbib}
\usepackage{amsmath,mathtools}
\usepackage{subfig}
\usepackage{graphicx}
\usepackage{url}
\usepackage[usenames,dvipsnames]{xcolor}
\usepackage{float}

\makeindex

\title[Monthly Notices: \LaTeXe\ guide for authors]
  {Chaotic dynamics of the planet in HD 196885 AB}
\author[Satyal et al.]
  {S.~Satyal,$^1$\thanks{e-mail: ssatyal@uta.edu}
  T.C.~Hinse$^{2,3}$ B.~Quarles,$^4$  J.P.~Noyola$^1$
  \\
  $^1$The Department of Physics, University of Texas at Arlington, Arlington, Texas, 76019, USA.\\
	$^2$Korea Astronomy and Space Science Institute, 304-358 Daejeon, Republic of Korea.\\
  $^3$Armagh Observatory, College Hill, BT61 9DG, Armagh, UK.\\
	$^4$Space Science and Astrobiology Division 245-3, NASA Ames Research Center, Moffett Field, CA 94035, USA.}
\date{Released 2014 January 6}

\pagerange{\pageref{firstpage}--\pageref{lastpage}} \pubyear{2002}

\def\LaTeX{L\kern-.36em\raise.3ex\hbox{a}\kern-.15em
    T\kern-.1667em\lower.7ex\hbox{E}\kern-.125emX}

\begin{document}

\label{firstpage}

\maketitle

\begin{abstract}
Depending on the planetary orbit around the host star(s), a planet could orbit either one or both stars in a binary system as S-type or P-type, respectively. We have analysed the dynamics of the S-type planetary system in HD 196885 AB with an emphasis on a planet with a higher orbital inclination relative to the binary plane. The mean exponential growth factor of nearby orbits (MEGNO) maps are used as an indicator to determine regions of periodicity and chaos for the various choices of the planet's semimajor axis, eccentricity and inclination with respect to the previously determined observational uncertainties. We have quantitatively mapped out the chaotic and quasi-periodic regions of the system's phase space which indicate a likely regime of the planet's inclination. In addition, we inspect the resonant angle to determine whether alternation between libration and circulation occurs as a consequence of Kozai oscillations, a probable mechanism that can drive the planetary orbit to a very large inclination. Also, we demonstrate the possible higher mass limit of the planet and improve upon the current dynamical model based on our analysis.

\end{abstract}


\section{Introduction}
The study of exoplanets has made a significant leap since the first planet around a solar type star, 51 Pegasi b \citep{may95}, was discovered almost two decades ago. The number of confirmed exoplanets has already surpassed most peoples' expectations with a count of 1792 as of June 2, 2014\footnote{\url{www.exoplanet.eu}}. New exoplanets are being added almost everyday into the database. The ultimate conquest of the contemporary exoplanet research is to find a planet within its respective habitable zone with a stable planetary orbit and a terrestrial mean density. Various ground based surveys and space based telescopes, such as the $Kepler$\footnote{\url{http://kepler.nasa.gov/}} mission and future missions such as the transiting exoplanet survey satellite (TESS) \citep{ric10} and the James Webb\footnote{\url{http://www.jwst.nasa.gov/}} space telescope are set for the detection and characterization of exoplanetary systems which will deliver a promising future in the search of terrestrial planets which could support life. Meanwhile, the search for exomoons is also an ongoing process and recent work by \cite{kip13} has produced a list of candidate hosts of such moons around transiting exoplanets. \cite{noy13} recently calculated the minimum required flux from the exoplanetary radio emissions in order to detect the exomoons using several ground based radio telescopes. Although an exomoon has yet to be detected, theorists have already determined relations to constrain an exomoon's habitability by considering energy flux (i.e., radiative and tidal) and orbital stability \citep{hel12,cun13}.

The focus of this paper, however, is the study of the dynamics of an exoplanet whose orbital eccentricity and inclination relative to the binary plane are significantly higher than the median of those discovered. The orbital stability of such a system depends on various orbital parameters and the long-term stability is vital for life to develop under current theories.  Thus, understanding the complete dynamics of a system bears vital requirement. Various numerical tools have been developed and used in the past to address the orbital configuration leading to the stability, instability or chaos of planets in a binary system. \cite{qua11} used the maximum Lyapunov exponent (MLE), originally developed by \cite{lya07}, to determine the orbital stability or instability for the circular restricted three-body problem (CRTBP) case. The mean exponential growth factor of nearby orbits (MEGNO) \citep{cin99} maps have been used to study the dynamical stability of irregular satellites \citep{hin10} and extrasolar planet dynamics \citep{goz01a,goz01b}. Recently, \cite{sat13} used MLE, MEGNO and the Hill stability time series methods \citep{sze08} to study the orbital perturbation of a planet due to the stellar companion in the binary systems of $\gamma$ Cephei and HD 196885.

Planets that are formed from a planetary disk are expected to have near circular orbits due to tidal circularization, yet roughly 19\% of known exoplanets with well-known orbital parameters have an eccentricity greater than 0.4\footnote{\url{www.exoplanet.eu}}. This estimate, however, could be biased and may not include the true population. The actual intrinsic phenomenon on how such highly eccentric orbits are formed while maintaining orbital stability is not quite fully understood, however some theories do exist that postulate such origins. \cite{koz62} and \cite{lid62} proposed a mechanism, now referred to as \emph{Lidov-Kozai} Mechanism, to explain the variations of a test particle's eccentricity and inclination. This mechanism was first applied to the exoplanets by \cite{hol97} while studying the chaotic variations in the eccentricity of the planet in 16 Cygni AB. Furthermore, its application to the planets includes the work done by \cite{inn97}, \cite{wu03}, \cite{fab07}, \cite{wu07}, \cite{ver10}, \cite{cor11}, \cite{lit11}, \cite{nao11}, \cite{kat11}, \cite{nao12}, \cite{nao13} and \cite{li14}. Some other mechanisms have also been proposed, such as planet-planet scattering to explain such variations (see for example, \cite{ras96,cha08,nag08}). In this work, we make use of the \emph{Lidov-Kozai} mechanism to explain such highly eccentric but stable orbits of the S-type planet, HD 196885 Ab. The planet is part of a binary star system (HD 196885 AB) whose eccentricity is observationally determined to be 0.48 and the planet's inclination is unconstrained which implies a value anywhere between zero and ninety degrees (for prograde motion) with the binary plane.

The dynamics of the planet in HD 196885 have been studied in a previous work by \cite{sat13}. In that study, investigations were performed in a restricted manner where the planet's inclination, (i$_{pl}$) $\le$ 25$^\circ$ with the binary plane, and the dynamics of the system was not fully explored. For this paper, we have considered a full range of prograde orbits in i$_{pl}$ value, from 0$^\circ$ to 90$^\circ$, and have made use of the chaos indicator, MEGNO, to produce maps which demonstrate regions of periodicity and chaos for a variety of initial conditions. For similar initial conditions a dynamical \emph{lifetime} map is produced by using the information about the planet's ejection \emph{time} from the system or the collision \emph{time} with the stellar host. Also, the dynamics of the system is studied in terms of the planet's maximum eccentricity. Then, the resonant angle is analyzed for evidence of a mean motion resonance at the best-fit location of the planet and for possible alternation between libration and circulation arising due to chaos induced by Kozai oscillations.

If a system demonstrates quasi-periodic orbits (as in the case of MEGNO maps) and/or the planet survives the total simulation time (\emph{lifetime} map), we consider it a stable system. If the planet gets ejected from the system or collides with the central body within the simulation time, it would be an unstable system. For some initial conditions the planetary orbit displays a chaotic motion, which we shall confirm from the evaluation of the time evolution of the resonant angle. Even though the motion is chaotic it does not mean that its unstable and the system can maintain (in some cases it has maintained) a stable configuration for millions of years.

This paper is outlined as follows. In Section 2 we present the basic theory of the \emph{Lidov-Kozai} mechanism, our numerical approach, a discussion of MEGNO as a chaos indicator and the initial set up of the system. In Section 3 we present our results from a representative sample of singly and multi integrated orbits and MEGNO maps followed by discussion. We conclude in Section 4 with a brief overview of our results.

\begin{table}
\begin{tabular}{|l|c|c|}
\hline
\textbf{HD 196885} & \textbf{B} & \textbf{Ab} \\
\hline
Mass (m)               & 0.45 M$_\odot$ & 2.98 M$_J$ \\
Semimajor Axis ($a$) & 21 $\pm$ 0.86 AU                & 2.6 $\pm$ 0.1 AU   \\
Eccentricity ($e$)   & 0.42 $\pm$ 0.03               & 0.48 $\pm$ 0.02     \\
Inclination ($i$)  & 0$^\circ$									& [0$^\circ$ - 90$^\circ$]	\\
Argument of Periapsis ($\omega_p$) & 241.9$^\circ$ $\pm$ 3.1    & 93.2$^\circ$ $\pm$ 3.0 \\
Ascending node ($\Omega$) & 79.8$^\circ$ $\pm$ 0.1      & 0.0$^\circ$ \\
Mean Anomaly ($M$)  & 121$^\circ$ $\pm$ 45    & 349.1$^\circ$ $\pm$ 1.8 \\
\hline
\end{tabular}
\caption{Best-fit orbital parameters of the HD 196885 system as obtained from \citep{cha11}. Mass of primary star, m$_A$ = 1.33 M$_\odot$. The planet's inclination ($i_{pl}$) is measured relative to the binary orbit,  then for $i_{pl}$ = 0$^\circ$, the planetary orbit is coplanar with the binary orbit.}
\label{tab:HD196885}
\end{table}

\section{Theory}
\subsection{The Lidov-Kozai Mechanism} \label{sec:KM}

		%

\cite{koz62} developed an analytical theory to explain the secular perturbations induced by Jupiter on the asteroids in the Solar System. Similar theory was developed by \cite{lid62} to study the evolution of orbits of artificial satellites of planets that are directly influenced by the gravitational perturbations of the Sun. For this, Kozai considered the perturber's (Jupiter) orbit to be circular. Thus the asteroid's vertical angular momentum and the secular energy is conserved and the system is integrable. As a consequence of the conservation of the quantity $\sqrt{1-e_{pl}^2}\cos{i_{pl}}$ the time evolution of eccentricity and inclination of the planetary orbit is anti-correlated: when the eccentricity is small, then the inclination is high and vice versa. For relative orbital inclinations less than 39.2 degrees the argument of pericentre circulates between 0$^\circ$ and 360$^\circ$ and the planet's orbit is precessing. This property remains true for other values of initial eccentricity. However, for relative inclinations larger than 39.2 degrees, the planet becomes a Kozai librator (Kozai regime) with it's argument of pericentre locked and exhibiting either librations (oscillations) around 90 or 270 degrees. These libration centers are known to be stable fixed points. A Kozai librator with an initially circular orbit will undergo a large variation in eccentricity and inclination within Kozai cycles. For relative inclinations close to or around the critical value of 39.2 degrees, the argument of pericentre exhibits intermittent behaviour displaying alternations between circulations and librations. In general, the coupling between eccentricity and inclination provides an effective removal mechanism. For a large initial relative inclination the amplitude of the planet's eccentricity variation increases. Following \cite{inn97} the maximum extent in eccentricity for a Kozai cycle can be expressed as,

\begin{align}
 (\emph{e}_{max})_{pl} &= {\sqrt{1-{5\over3}\cos^2{(\emph{i}_{o})_{pl}}}},
\label{eqn:kozai}
\end{align}

where $(\emph{i}_{o})_{pl}$ is the initial relative inclination of the planet. Eventually, for a large enough eccentricity, the planet is either ejected from the system or collides with the primary component rendering its orbit to be unstable.

However, the vertical angular momentum and the secular energy are conserved only when the pertuber's orbit is circular. When the perturber is eccentric or if the planet has non-negligible mass, the vertical angular momentum component of the planet and the perturber is not conserved and the planet shows qualitatively different behaviour. For such cases, \cite{lit11} and \cite{nao13} have further developed the \emph{Lidov-Kozai} mechanism discussed earlier and formulated for the eccentric perturber, which they have called it the \emph{eccentric Kozai mechanism} (EKM). When the perturber has an eccentric orbit the quadrupole-order approximation is insufficient and requires the octupole-order term as well \citep{for00}. Following \cite{lit11} and \cite{nao13}, the energy function and the constant of motion, F, in quadrupole and octupole term is, \emph{F} $\equiv$ $F_{quad}$ + $\epsilon$$F_{oct}$ and the constant is given by

\begin{align}
 \epsilon \equiv {a_{pl}\over a_{per}} {e_{per} \over{1 - e_{per}^2}},
\label{eqn:EKM}
\end{align}

where $a_{pl}$ and $a_{per}$ are the semimajor axis of a planet and pertuber respectively, and $e_{per}$ is the pertuber's eccentricity. In a binary star system with an S-type planet around the primary star, as in our case, the secondary star would be a perturber. When $e_{per}$ = 0, $\epsilon$ = 0, thus F reduces to the case with quadrupole term only. Further mathematical set up of equations of motion concerning the quadrupole and octupole order terms can be found in \cite{nao13}. During their study of the EKM, some of the remarkable results were obtained including the flipping of the planetary orbit from prograde to retrograde and its eccentricity reaching to the extreme values ($\sim$ 1). Two main cases have been discussed regarding the initial conditions in $e_{pl}$ and $i_{pl}$: high eccentricity, low inclination (HeLi) and low eccentricity, high inclination (LeHi). Primarily, only the octupole term is in play during HeLi flip \citep{li14} and both terms (quadrupole and octupole) are in play during LeHi flip \citep{nao11,nao13}. From \cite{li14}, the flip criterion is given as,

\begin{align}
 \epsilon > {{8\over5} {{1 - e_{per}^2} \over{7 - e_{per}(4+3e_{per}^2)cos(\omega_{per} + \Omega_{per})}}},
\label{eqn:flip_criterion}
\end{align}

where $\epsilon$ is the constant parameter (Eq. \ref{eqn:EKM}), $e_{per}$, $\omega_{per}$ and $\Omega_{per}$ are the perturber's initial eccentricity, argument of periapsis and ascending node, respectively.

		%

\subsection{Numerical Approach and methods}

\begin{figure*}
\centering
\subfloat[Initial $i_{pl}$ = 0$^\circ$]{\includegraphics[width=.33\linewidth]{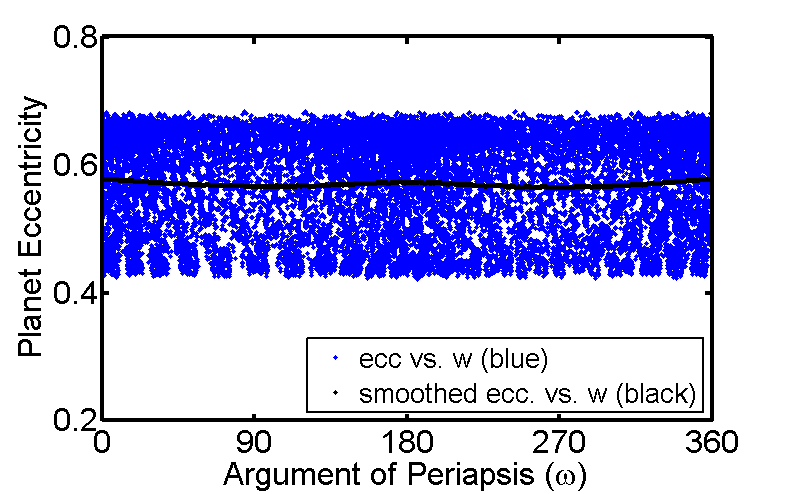}}
\subfloat[Initial $i_{pl}$ = 20$^\circ$]{\includegraphics[width=.33\linewidth]{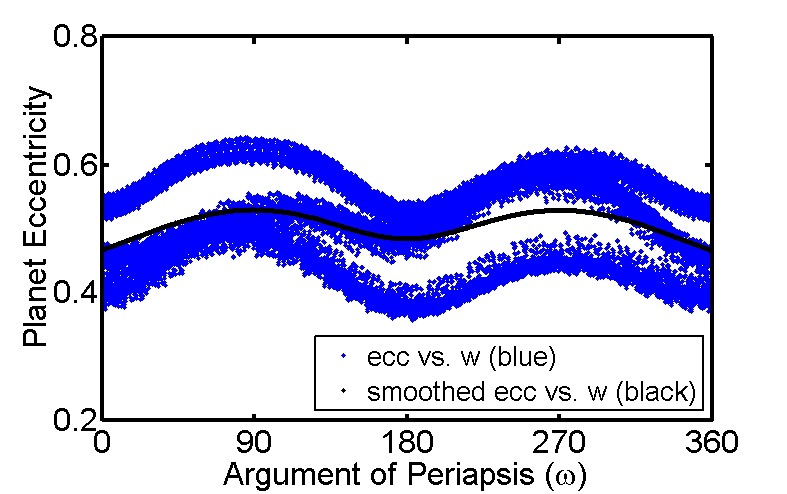}}
\subfloat[Initial $i_{pl}$ = 35$^\circ$]{\includegraphics[width=.33\linewidth]{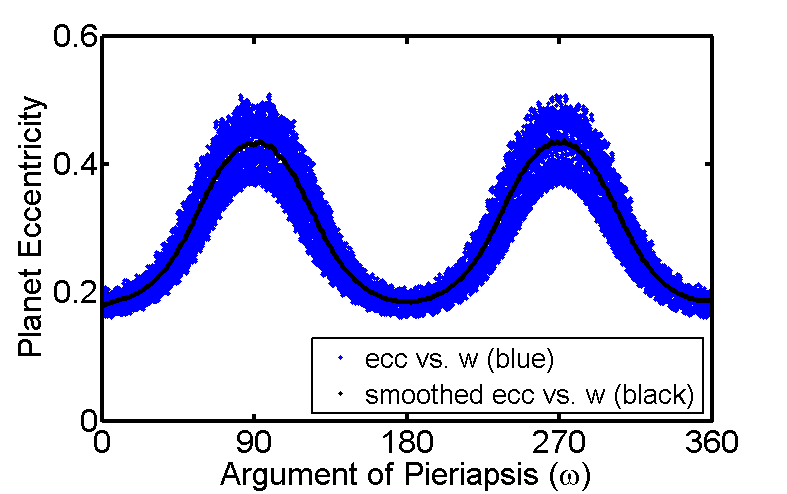}}
\\*
\subfloat[Initial $i_{pl}$ = 39.2$^\circ$]{\includegraphics[width=.33\linewidth]{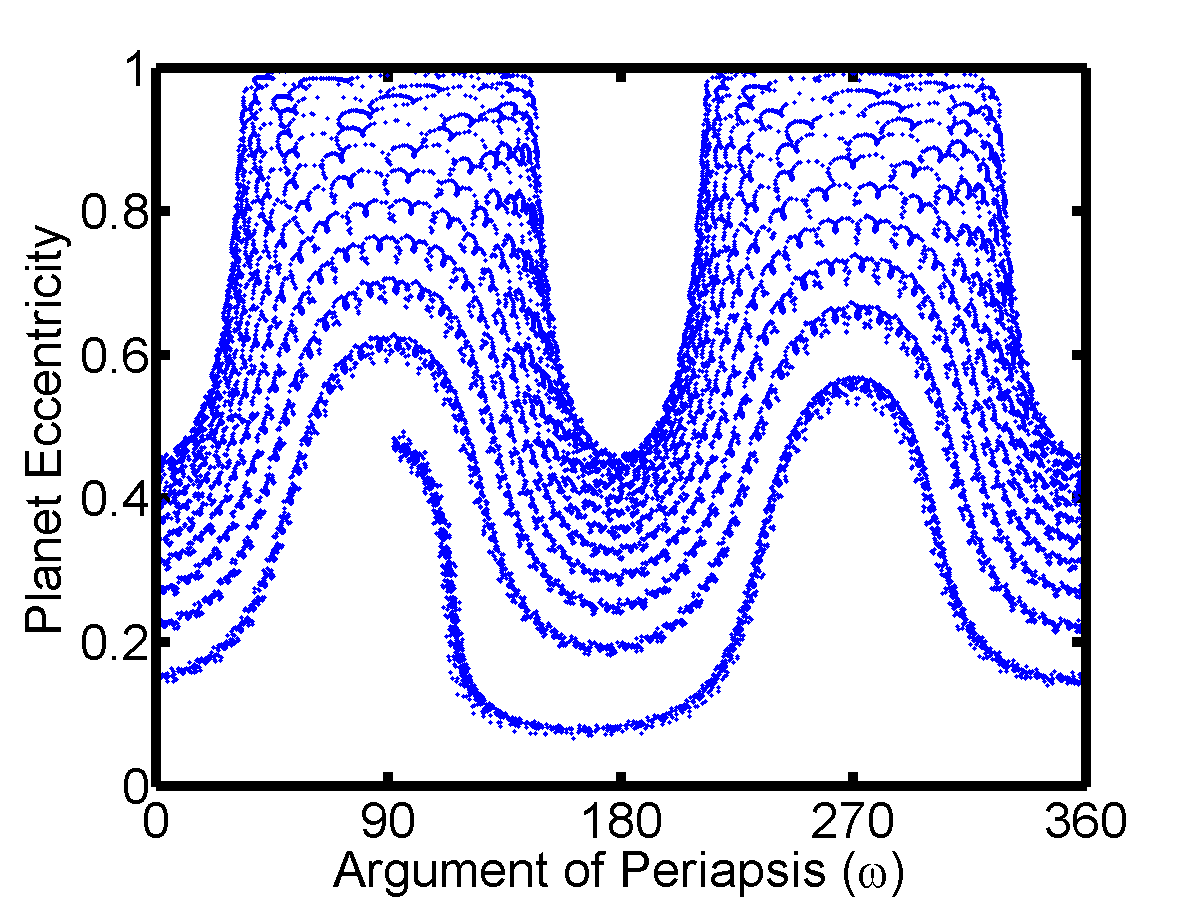}}
\subfloat[Initial $i_{pl}$ = 39.7$^\circ$]{\includegraphics[width=.33\linewidth]{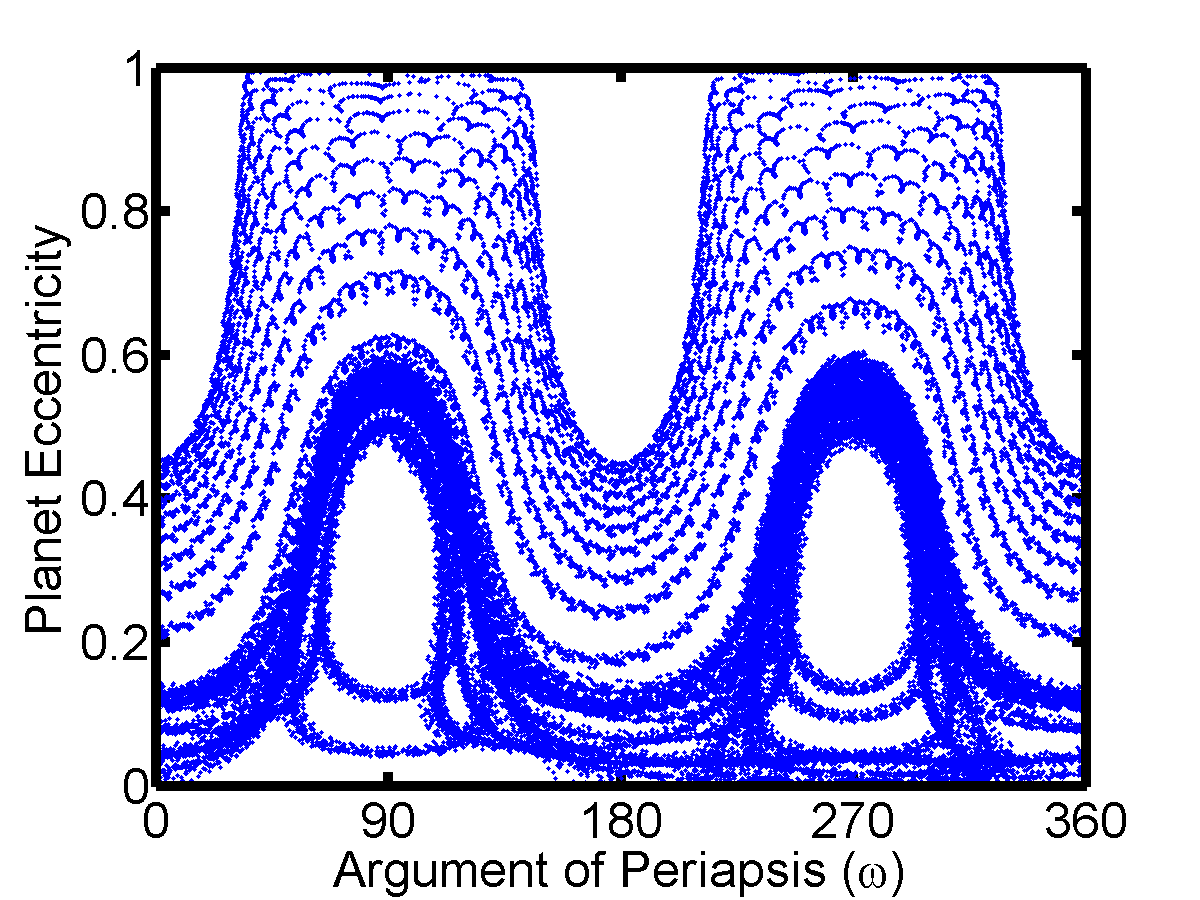}}
\subfloat[Initial $i_{pl}$ = 45$^\circ$, 60$^\circ$, 70$^\circ$, 80$^\circ$]{\includegraphics[width=.33\linewidth]{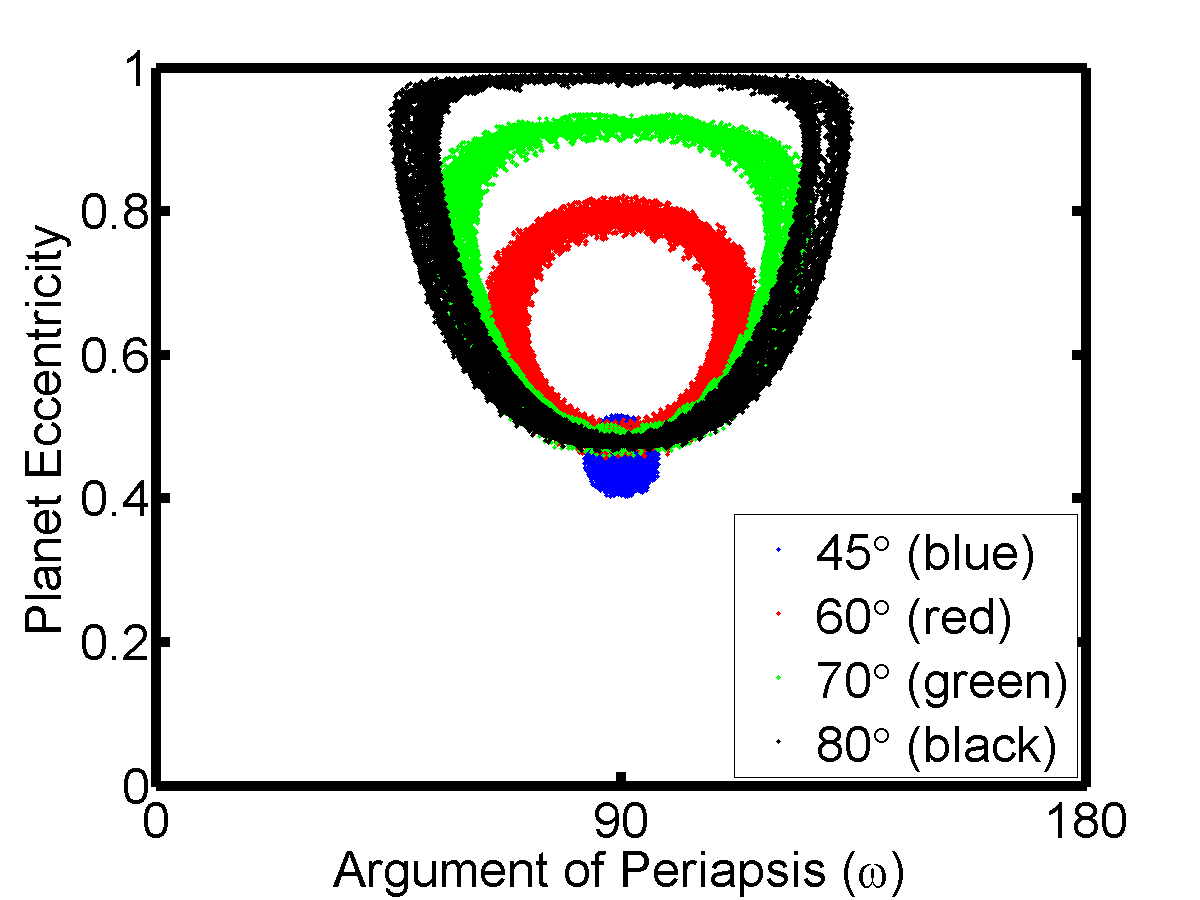}}\
\caption{Evolution of planet's eccentricity and argument of periapsis for various choices of planet's orbital inclination with the binary plane, simulated for 1 $\times 10^5$ years. The Savitzky-Golay smoothing function with third degree polynomial is used to smooth the high frequency data as shown in (a),(b) and (c). For the first three choices of $i_{pl}$ values, the $\omega$ is circulating within 0$^\circ$ to 360$^\circ$. In contrast to the coplanar case, the amplitude of circulation is varying and is maximum at 90$^\circ$ and 270$^\circ$ (b and c). When $i_{pl}$ = 39.2$^\circ$ (d), the $\omega$ shows circulation in the region of phase space with eccentricity amplitude maximum at 90$^\circ$ and 270$^\circ$, and for even higher $i_{pl}$ (39.7$^\circ$) the phase space divides into two distinct regions where $\omega$ shows circulation and libration around 90$^\circ$ and 270$^\circ$ (e). This is, however, an unstable region where the amplitude of eccentricity oscillation reaches $\sim$1. When $i_{pl}$ is set to 45$^\circ$, 60$^\circ$, 70$^\circ$ and 80$^\circ$ (f), the $\omega$ librates around 90$^\circ$ and the libration amplitude in eccentricity is found to increase with increasing $i_{pl}$ until the planet collides with the central body for $i_{pl}$ greater than 80$^\circ$.}
\label{fig:ew}
\end{figure*}

Our approach has considered the motion of a planet of mass, $m_{pl}$ around a star of mass, m$_A$. When calculating the initial conditions in position and velocity, we have used the best-fit orbital elements, semi major axis ($a$), eccentricity ($e$), inclination ($i$), argument of periapsis ($\omega$), ascending node ($\Omega$) and mean anomaly ($M$), that are obtained from the radial velocity measurements \citep{cha11}. For any unconstrained parameters, the values are taken in a range or considered zero. The list of orbital parameters of the HD 196885 system are given in Table \ref{tab:HD196885}. The subscripts \emph{bin} and \emph{pl} are used to denote the secondary binary component and the planet, respectively. The mean longitude ($\lambda$) that is used to calculate the resonant angle ($\Phi$) is calculated from the longitude of the periapsis ($\varpi$ = $\Omega + \omega$) and the orbit's mean anomaly ($M$), which is given as $\lambda$ = $\varpi$ + $M$.

Using the orbital integration package MERCURY \citep{cha97,cha99}, the built-in Radau algorithm was used to integrate the orbits of the system in astrocentric coordinates when investigating the evolution of orbital elements for a single initial condition and to produce a \emph{lifetime} map for multiple initial conditions. MERCURY was effective in monitoring the ejection/collision of a planet due to close encounter with the secondary star and provided robust results for the purpose of determining the existence of orbital resonance. A time step of $\epsilon = 10^{-3}$ year/step was considered to have high precision because the error in change in total energy and total angular momentum was calculated at each time step which fell within the range of 10$^{-16}$ to 10$^{-13}$ in both cases during the total integration period of 50 Myr. The data sampling (DSP) was done per year for shorter integration periods (up to 100 kyr) and per thousand years for longer integration periods. The \emph{lifetime} map and the maximum eccentricity map are generated for 12,000 initial conditions in $a_{pl}$ and $i_{pl}$, and simulated for 50 Myr.

We used the MECHANIC\footnote{https://github.com/mslonina/Mechanic} software \citep{slo12,slo14} optimised to $N$-body code to calculate the orbits of the given masses and the MEGNO maps on a multi-CPU computing environment. The MEGNO criterion is used to differentiate between the periodic and chaotic dynamics of a system. Originally developed by \cite{cin99, cin00}, MEGNO has been widely used in the study of dynamical astronomy \citep[and references therein]{goz01b,goz01a,hin08,hin10}. For reason of completeness we outline some details of the MEGNO formalism. Following \cite{cin00} MEGNO ($Y$) is defined as

\begin{equation}
Y(t) =
\frac{2}{t}\int_{t_{0}}^t\frac{\parallel\dot{\boldsymbol\delta}(s)\parallel}
{\parallel\boldsymbol\delta(s)\parallel}~s~ds,
\label{MEGNO}
\end{equation}
along with its time-averaged mean value
\begin{equation}
\langle Y\rangle(t) = \frac{1}{t}\int_{t_{0}}^tY(s)~ds.
\label{time-averageMEGNO}
\end{equation}
\noindent
The quantity $\boldsymbol\delta$ is the variational state vector and $t$ is time. The absolute norm of $\boldsymbol\delta$ measures the distance between 
two nearby points in phase space using a euclidian metric. The variational vector is found from solving the variational equations of motion in parallel to the newtonian equations of motion \citep{mik99,goz01b,hin10}. Equations \ref{MEGNO} and \ref{time-averageMEGNO} can be rewritten \citep{goz01b} as two first-order differential equations
\begin{equation}
\frac{dx}{dt} = \frac{\dot{\boldsymbol{\delta}}\cdot {\boldsymbol{\delta}}}
{\parallel\boldsymbol{\delta}\parallel^2}~t~~~~~
\textnormal{and}~~~~~\frac{dw}{dt} = 2~\frac{x}{t},
\end{equation}
\noindent
and are solved in tandem alongside the equations of motion and variational equations of motion. At the end of each integration step the quantities 
$Y(t)$ and $\langle Y \rangle (t)$ can be determined from $Y(t) = 2x(t)/t$ 
and $\langle Y\rangle(t)=w(t)/t$. When generating MEGNO maps we chose to plot the mean value $\langle Y \rangle$. 

In general, MEGNO has the parameterisation $\langle Y \rangle = \alpha \times t + \beta$ \citep{cin00,goz01b}. For a quasi-periodic initial condition, we have $\alpha \simeq 0.0$ and $\beta \simeq 2.0$ (or $\langle Y \rangle \rightarrow 2.0$) for $t \rightarrow \infty$ asymptotically. If the orbit is chaotic, then $\langle Y\rangle \rightarrow \lambda t/2$ for $t \rightarrow \infty$. Here $\lambda$ is the maximum Lyapunov exponent (MLE) of the orbit. In practice, when generating our MEGNO maps, we terminate a given numerical integration of a chaotic orbit when $\langle Y \rangle > 12.0$. In contrast, quasi-periodic orbits have $|\langle Y\rangle - 2.0| \le 0.001$. For more details on the mathematical properties of MEGNO and its relationship with Lyapunov exponents, see \cite[and references therein]{hin10}.

In our study, each map is generated with 1.5$\times10^5$ initial conditions, for a resolution of ($300 \times 500$) in various orbital elements and simulated up to 200 Kyr. The purple/blue color in the map denotes the quasi periodic region and the yellow is the region of chaos. The MEGNO quantity $\langle Y\rangle$ is colour-coded in the colour bar.


\section{Results and Discussion}\label{sec:results}
\subsection{The Phase Space Evolution}

\begin{figure*}
\centering
\hspace*{\fill}%
\subfloat[Initial $i_{pl}$ = 10$^\circ$]{\includegraphics[width=.33\linewidth]{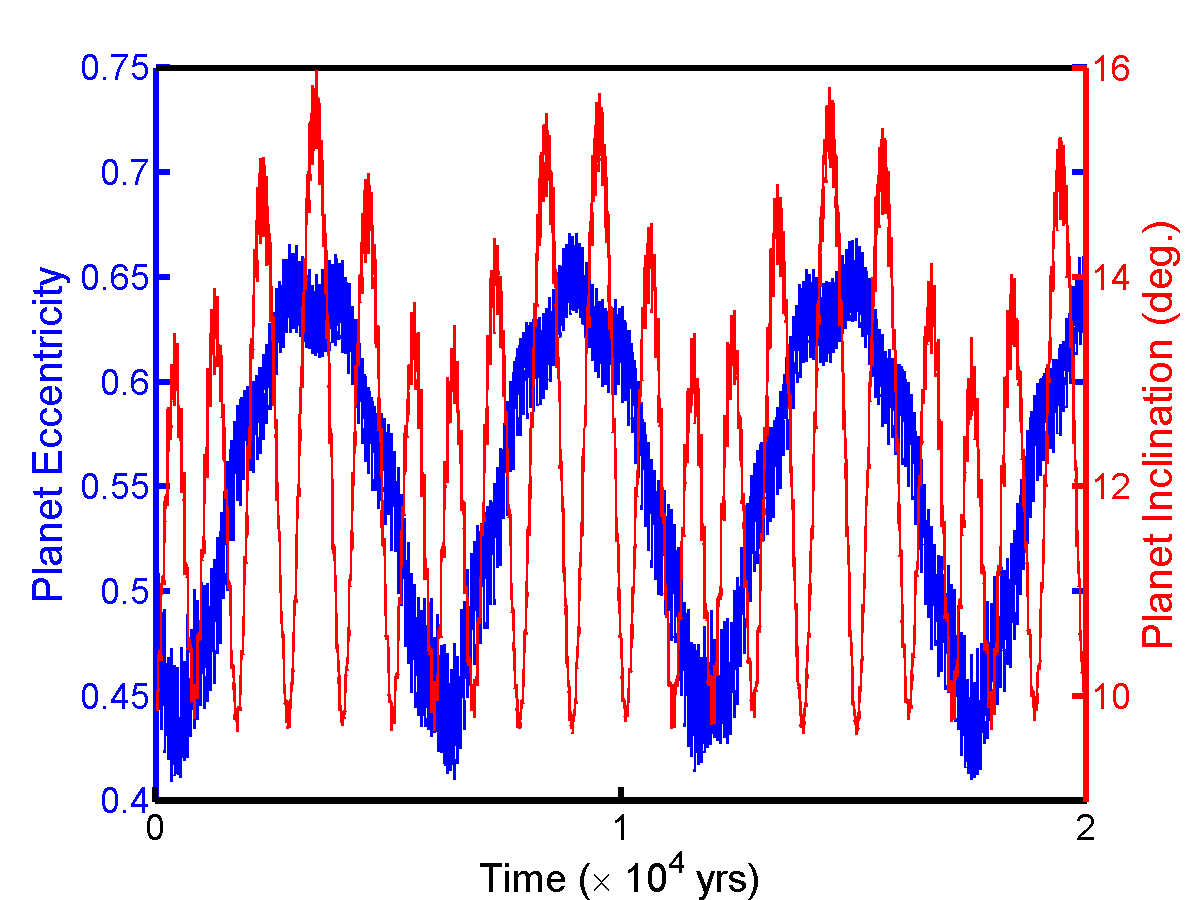}}\hfill
\subfloat[Initial $i_{pl}$ = 30$^\circ$]{\includegraphics[width=.33\linewidth]{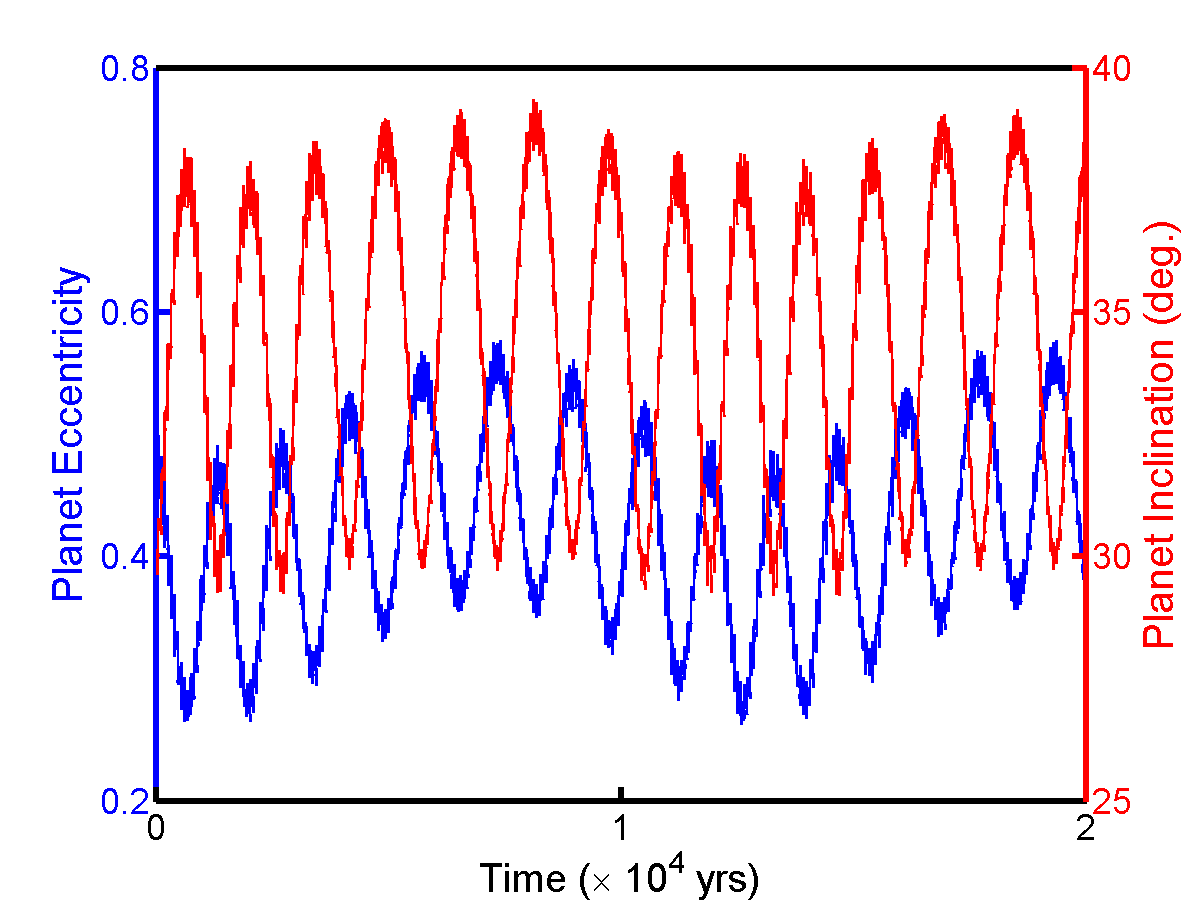}}\hfill
\subfloat[Initial $i_{pl}$ = 39.2$^\circ$]{\includegraphics[width=.33\linewidth]{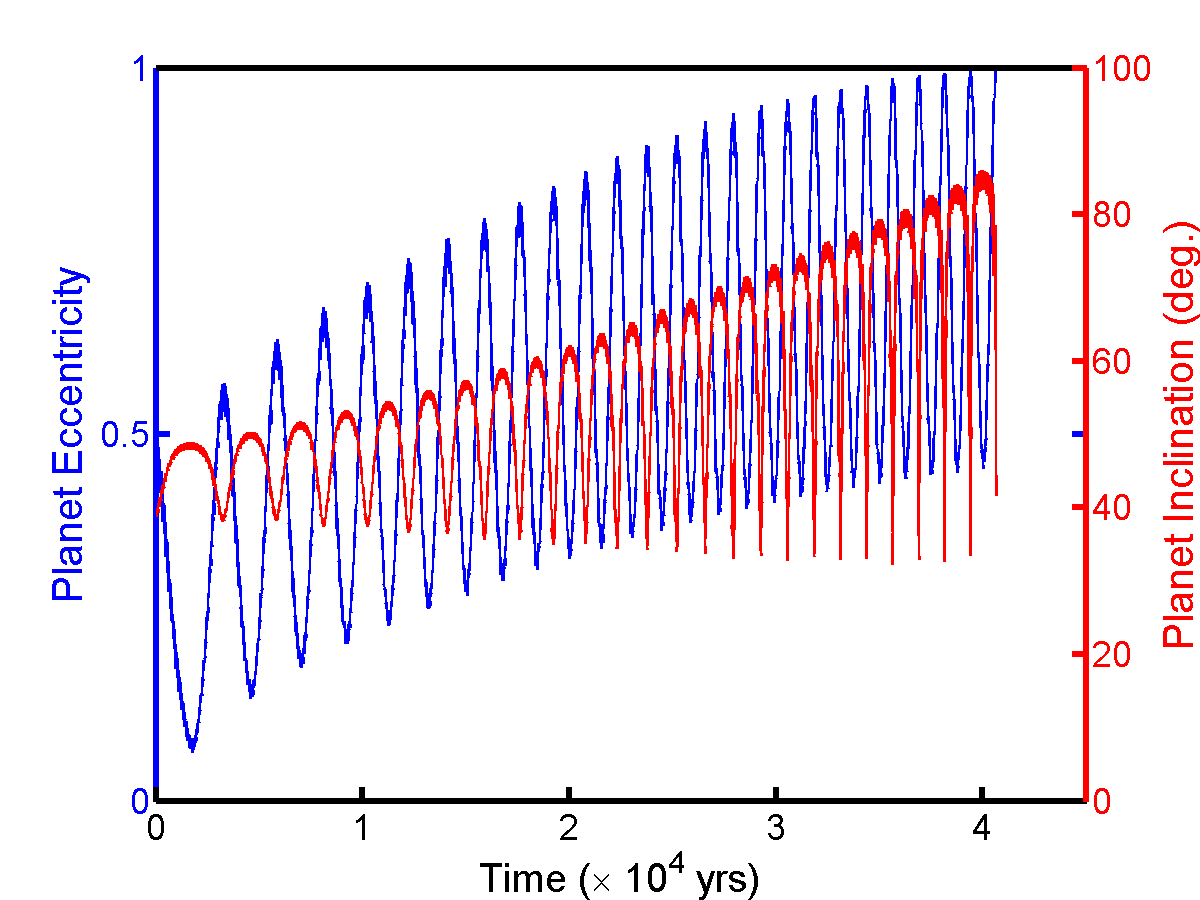}}
\hspace*{\fill}%
\caption{Planet's eccentricity and inclination time series when planet's initial orbital inclination with the binary plane is set at 10$^\circ$, 30$^\circ$ and 39.2$^\circ$. In these double-axis plots, the left \emph{y}-axis has eccentricity and right \emph{y}-axis has inclination plotted versus time along the common \emph{x}-axis. The orbital integration was carried out for 50 Myr but (a) and (b) are truncated at 2$\times$10$^4$ years to clearly demonstrate the oscillations of both elements (\emph{e} and \emph{i}). (c) is plotted up to the instability point ($\sim$ 4$\times$10$^4$ years) after which the planet collided with the central body.}
\label{fig:ecc_incl1}
\end{figure*}

\begin{figure*}
\centering
\hspace*{\fill}%
\subfloat[Initial $i_{pl}$ = 50$^\circ$]{\includegraphics[width=.33\linewidth]{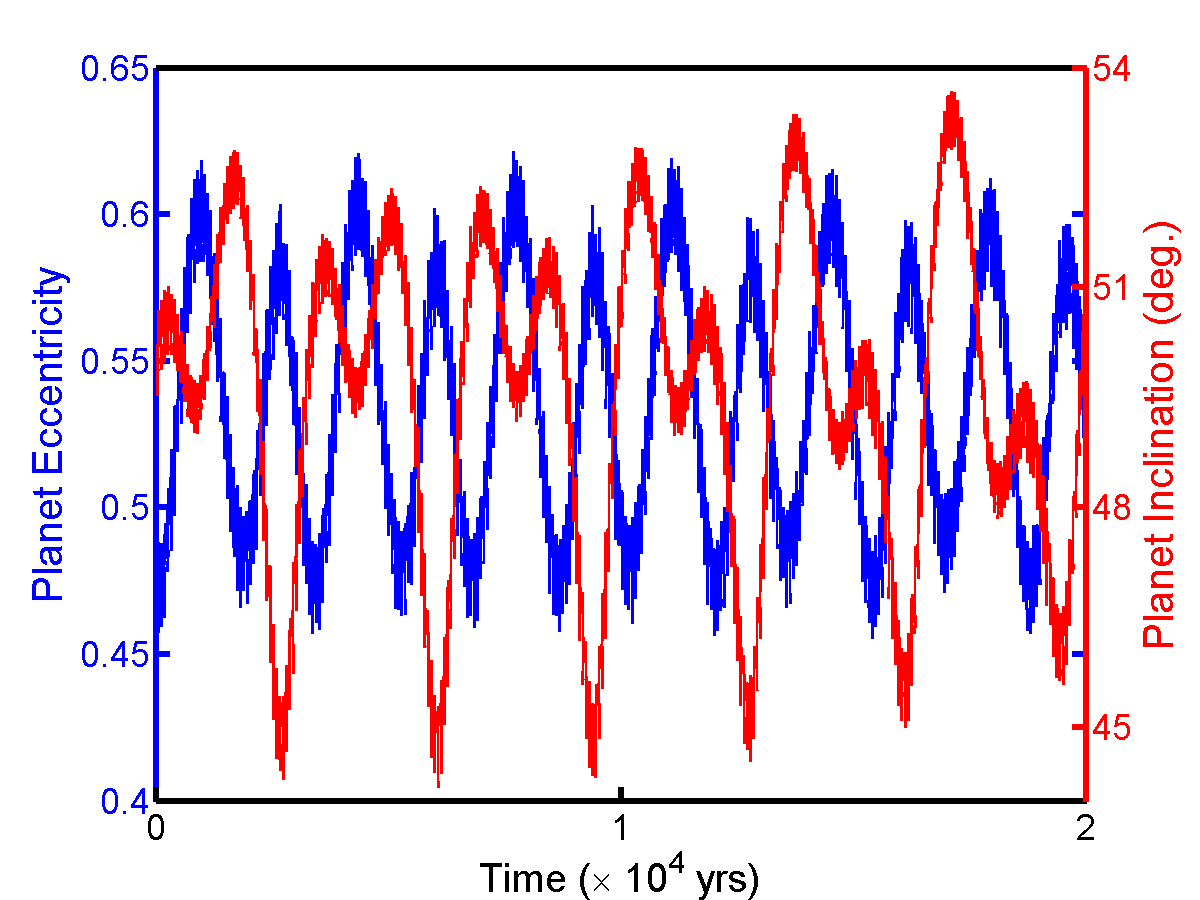}}\hfill
\subfloat[Initial $i_{pl}$ = 60$^\circ$]{\includegraphics[width=.33\linewidth]{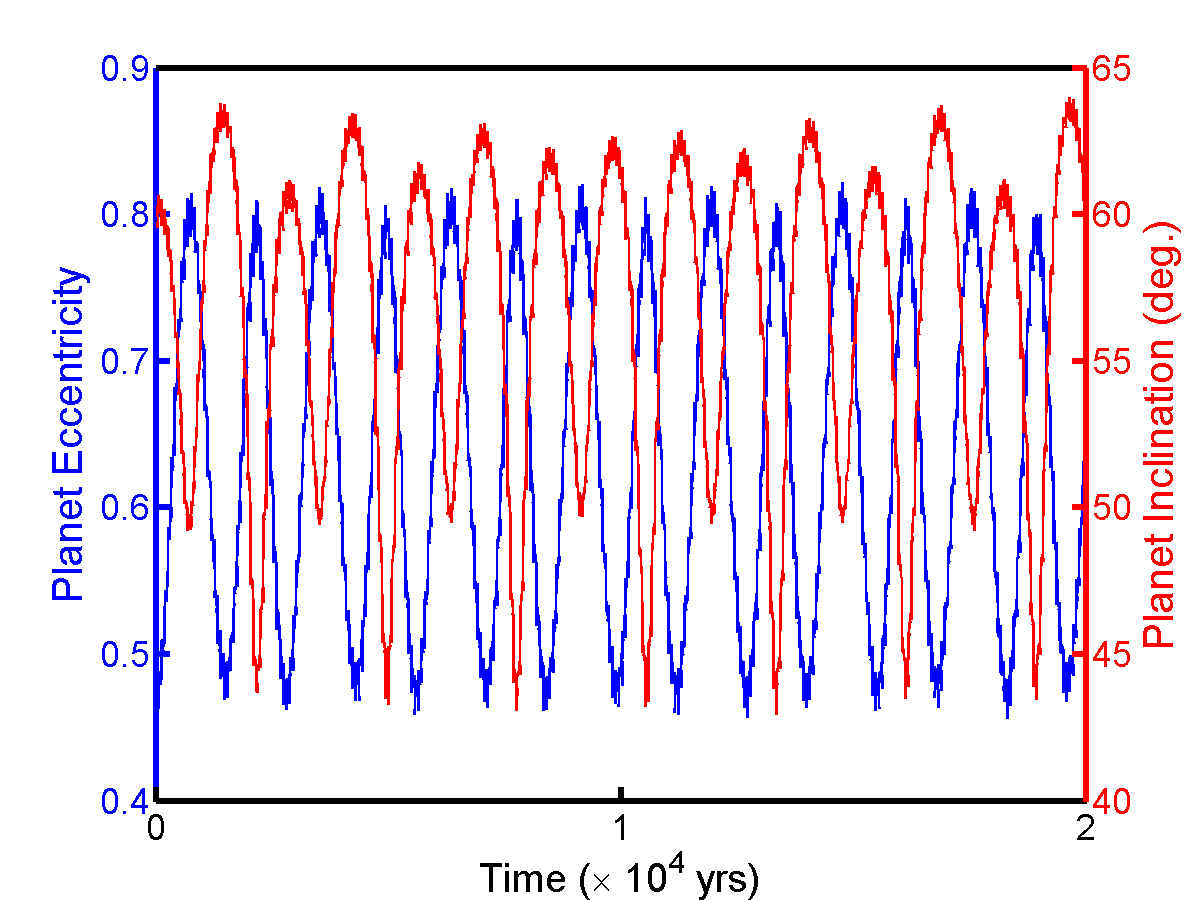}}\hfill
\subfloat[Initial $i_{pl}$ = 83$^\circ$]{\includegraphics[width=.33\linewidth]{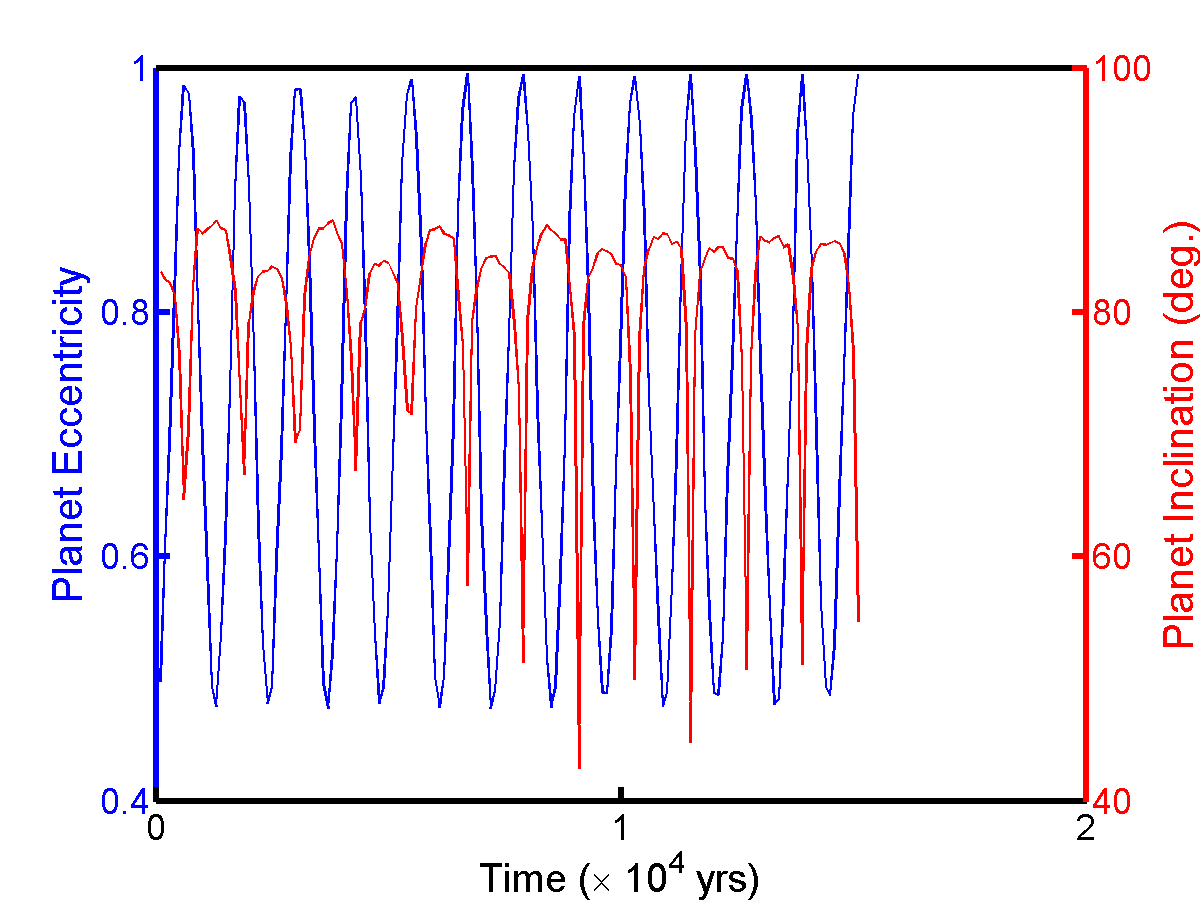}}
\hspace*{\fill}%
\caption{Planet's eccentricity and inclination time series when planet's initial orbital inclination with the binary plane is set at 50$^\circ$, 60$^\circ$ and 83$^\circ$. The time series at (a) and (b) are truncated at 2$\times$10$^4$ years and the time series at (c) is plotted up to the instability point ($\sim$ 1.5$\times$10$^4$ years) after which the planet collided with the central body.}
\label{fig:ecc_incl2}
\end{figure*}

To observe the system that is evolving in the phase space, the planet's eccentricity ($e$) is plotted versus the argument of periapsis ($\omega$), for various choices of the planet's relative orbital inclination with the binary plane (Fig. \ref{fig:ew}). The Savitzky-Golay smoothing function is used to filter the data as we are interested in the secular dynamics of the system and the secular perturbation theory which initiates the Kozai mechanism in the case of inclined orbits. We find that, below 39$^\circ$ and for our choice of $i_{pl}$ values, the phase space starts to evolve where $\omega$ starts to circulate between 0$^\circ$ and 360$^\circ$. The eccentricity, however, oscillates with large and increasing amplitude for increasing $i_{pl}$ (for example 20$^\circ$ to 35$^\circ$) while compared to the coplanar case (see Figs. \ref{fig:ew}a,b,c).

When the initial $i_{pl}$ is set at 39.2$^\circ$, the $\omega$ shows circulation only with eccentricity amplitude maximum at 90$^\circ$ and 270$^\circ$ (Fig. \ref{fig:ew}d), but the system becomes orbitally unstable as the $e_{pl}$ reaches its maximum value causing the planet to collide with the central body within 50 Kyr. Further increment in the $i_{pl}$ value to 39.7$^\circ$ makes the phase space divide into two distinct regions where the $\omega$ is found to circulate as well as librate (Fig. \ref{fig:ew}e). This is also a transition regime of the phase space beyond which the $\omega$ displays libration only. \cite{hol97} have shown a similar behaviour in the momemtum variable (1-$e^2$) plotted versus its conjugate angle ($\omega$) for a highly inclined ($i_{pl}$ = 60$^\circ$) planetary orbit.

Furthermore, when the $i_{pl}$ is set in increasing order from 45$^\circ$ to 80$^\circ$, the amplitudes of the $e_{pl}$ oscillations is found to increase as well, and in this inclination regime the argument of periapsis only shows libration around 90$^\circ$ (Fig. \ref{fig:ew}f). The libration amplitude increases with the increase in the initial $i_{pl}$ values, eventually entering a regime where the planet can escape from the system or collide with its host star when $i_{pl}$ reaches a value greater than 82$^\circ$ and $e_{max}$ approaches 1.0 (Eq. \ref{eqn:kozai}).

\subsection{The Kozai Resonance}\label{sec:KM}

The evolution of a planet's eccentricity and inclination was further explored for a wide range of $i_{pl}$ values. The conservation of the Kozai integral term (see Sec. \ref{sec:KM}) suggests that as the system evolves in time, the decrease/increase in planet's eccentricity is compensated by the increase/decrease in it's inclination relative to the binary plane, also shown by \cite{hol97} for circular perturber and by \cite{lit11} for an eccentric perturber. For small initial $i_{pl}$ values (i.e., 10$^\circ$) we found a constant-amplitude oscillation of $e_{pl}$ and $i_{pl}$ at the average values of 0.55 and 12$^\circ$, respectively (Fig. \ref{fig:ecc_incl1}a) for the integration period of 50 Myr. As the initial value of $i_{pl}$ is increased the amplitude of oscillations of both parameters increase as well. In Fig. \ref{fig:ecc_incl1}b we have shown a case for $i_{pl}$ = 30$^\circ$. In this double-axis plot, the \emph{y}-axis on the left side indicates the time evolution of eccentricity from its best-fit value of 0.48. As the system evolves, the  eccentricity varies between 0.25 and 0.55. The \emph{y}-axis on right side illustrates the variation in inclination from its initial assigned value of 30$^\circ$. The $i_{pl}$ value oscillates between 30$^\circ$ and 40$^\circ$ and the variation in its amplitude from the initially assigned value behaves in an anti-correlated fashion to that of $e_{pl}$. Also, the variation in the $e_{pl}$ agrees with the case when the perturber has an eccentric orbit, the inner orbit's eccentricity can reach extremely large values \citep{for00, nao13, tey13}. The x-axis in Figs. \ref{fig:ecc_incl1}a and \ref{fig:ecc_incl1}b are truncated at 20 Kyr to clearly demonstrate the oscillations of the two parameters, ($e_{pl}$ and $i_{pl}$). The amplitude of the oscillations merely changed during its evolution for given total integration time.

When the initial $i_{pl}$ value was further increased to 39.2$^\circ$, the planet follows a path to instability within 40 kyr through a growing value of eccentricity.  It is also found that there is a small instability window (most of the time the planet collided with the central body) when $i_{pl}$ is set at $\sim$(39$^\circ$ - 40$^\circ$). The time series plot of $e_{pl}$ and $i_{pl}$ (Fig. \ref{fig:ecc_incl1}c) shows increasing values of the planet's eccentricity and inclination until it reaches an extreme value $\sim$1 and $i_{pl}$ oscillates up to 80$^\circ$, hence minimizing the periastron distance and eventually forcing the planet to collide with its host star. From a theoretical aspect it can be inferred that when the relative inclination hits the critical angle mark (\emph{i}$_c$ = 39.2$^\circ$), the long-period oscillations between eccentricity and inclination ensue. The initial eccentricity becomes insensitive leading to forced eccentricity, which is the basis for the Kozai resonance. The other factor causing the planet to exhibit Kozai cycles is the libration of the argument of periapsis ($\omega$) around 90$^\circ$, whose significant evolution is observed in Fig. \ref{fig:ew} for $i_{pl}$ greater than 39$^\circ$. We have not considered the tidal dissipation within the regimes of close approach which may as well produce misalignment \citep{win10}. Also, general relativity is not included in our calculations the precession of which can lead to large eccentricity oscillations \citep{nao13a,ant13}.

For greater orbital inclination values, such as, when initial $i_{pl}$ is set to 50$^\circ$ (or, 60$^\circ$) (Fig. \ref{fig:ecc_incl2}a,b) the amplitudes of eccentricity and inclination oscillations reach as high as $\sim$ 0.6 and 53$^\circ$ (or, $\sim$ 0.8 and 64$^\circ$), respectively, and the system continues to maintain the periodic orbits throughout the integration period of 50 Myr. Also, further increasing the $i_{pl}$ value pushes $e_{pl}$ to its extreme limit resulting in the planet-star collision when the $i_{pl}$ value is increased beyond 82$^\circ$. Above the critical value of the planet's orbital inclination with the binary plane ($i_c$ = 39.2$^\circ$), the precession of the argument of pericentre is replaced by libration around 90$^\circ$ as discussed earlier. \cite{sat13} encountered a similar dynamical behaviour in their study of the planet in the HD 196885 system for the planet's orbital inclination less than 25 degrees. For higher initial $i_{pl}$, as shown here, the secular perturbation causes the $e_{pl}$ and $i_{pl}$ values to oscillate with higher amplitudes, which can eventually cause instability in the system due to the collision of the planet with the central body or escape due to a parabolic or hyperbolic orbit. For example, when the initial $i_{pl}$ = 83$^\circ$ and the $e_{pl}$ is forced beyond 1 within a short integration period (15 kyr, Fig. \ref{fig:ecc_incl2}c).

\begin{figure}
\centering
\subfloat{\includegraphics[width=1\linewidth]{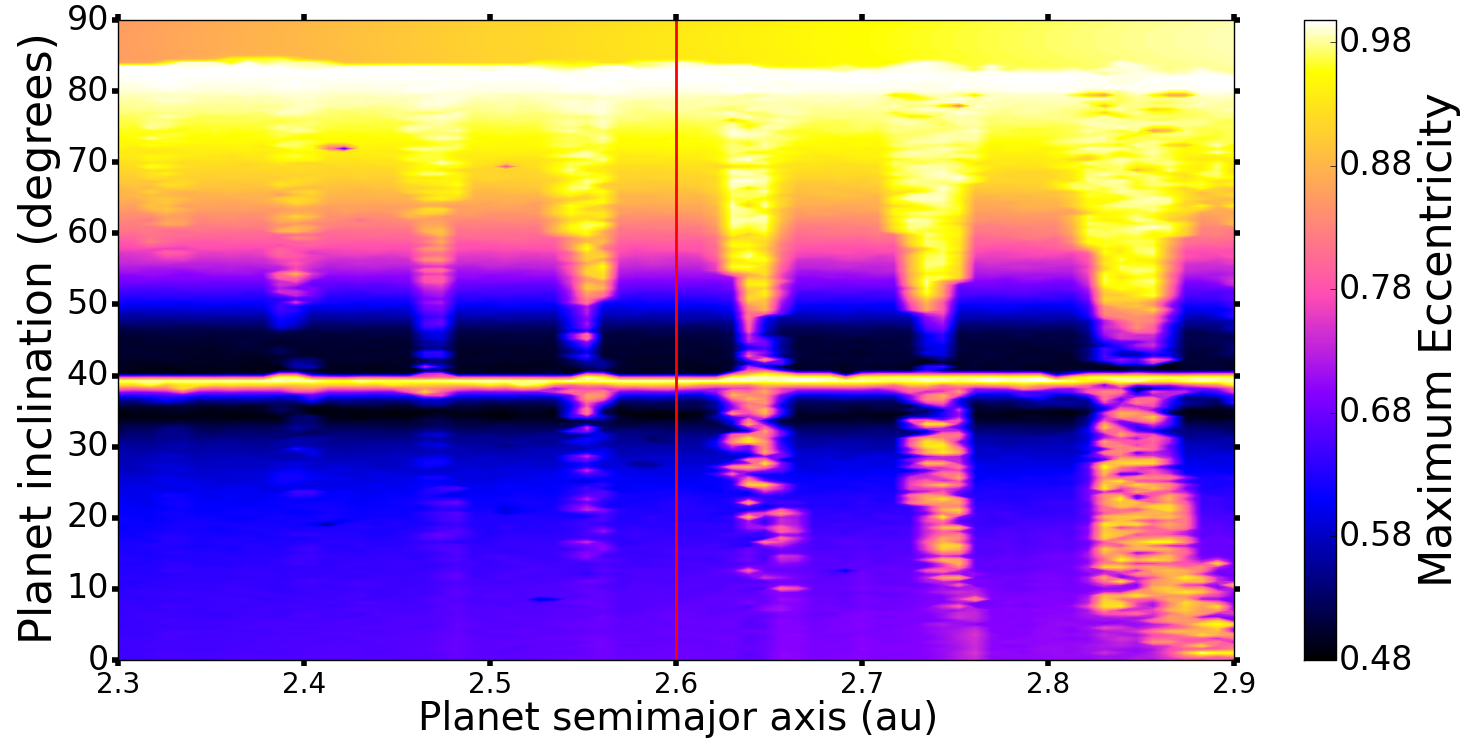}}
\caption{A maximum eccentricity ($e_{max}$) map of the planet, HD 196885 Ab, for varying $i_{pl}$ and $a_{pl}$, simulated for 50 Myr. The colour bar indicates the $e_{max}$ reached by the planet during the total simulation time, which also includes the cases when the planet suffers an ejection or collision (especially when $e_{pl}$ reaches a value greater than 0.9). The darker colour represents the best-fit $e_{pl}$ parameter (0.48) and the lightest colour represents the $e_{max}$ value the planet attained for the respective choices of initial conditions in $i_{pl}$ and $a_{pl}$. The red line at 2.6 au is the best-fit semimajor axis of the planet. This map compliments the \emph{lifetime} map (Fig. \ref{fig:stability}) in a way that the vertical instability bars are located at the same regions where the planet's $e_{max}$ exceeds 0.7.}
\label{fig:emax}
\end{figure}

For specific initial conditions, \cite{lit11} have shown that the planetary orbit can flip when the criterion given in Eqn. \ref{eqn:flip_criterion} is satisfied. But the initial parameters from our system does not satisfy this criterion despite the presence of the highly eccentric perturber. For the given initial parameters of HD 196885 AB, $\epsilon$ = 0.063 ($\epsilon$ is a constant term given by Eqn. \ref{eqn:EKM}) while the value obtained from  Eqn. \ref{eqn:flip_criterion} is 0.174. Thus, the planet is not expected to flip its orbit which is in agreement with Fig. \ref{fig:ecc_incl2}c and Fig. \ref{fig:emax}. The planet's maximum eccentricity reaches the extreme values, from 0.7 to $\sim$ 1, (see the maximum eccentricity map, Fig. \ref{fig:emax}) for the orbital inclination greater than 55$^\circ$; nonetheless, the planet survives the total integration time until the $i_{pl}$ is raised to 82$^\circ$ . The $e_{max}$ stays relatively low, between 0.48 and 0.68, for the $i_{pl}$ values less than 55$^\circ$ excluding the case when $i_{pl}$ is $\sim$ 39$^\circ$. Before reaching the flipping point (90$^\circ$) the planet gets ejected from the system when $i_{pl}$ $\ge$ 82$^\circ$ and the $e_{max}$ reaches close to 1.0 within the integration time.  We note that $e_{max}$ appears to decrease in Fig. \ref{fig:emax} for very high inclinations (85$^\circ-$90$^\circ$), but this is due to the approach to instability occurring faster than our data output frequency.

\subsection{The MEGNO and the MMR Analysis}

\begin{figure*}
\centering
\subfloat{\includegraphics[width=.48\linewidth]{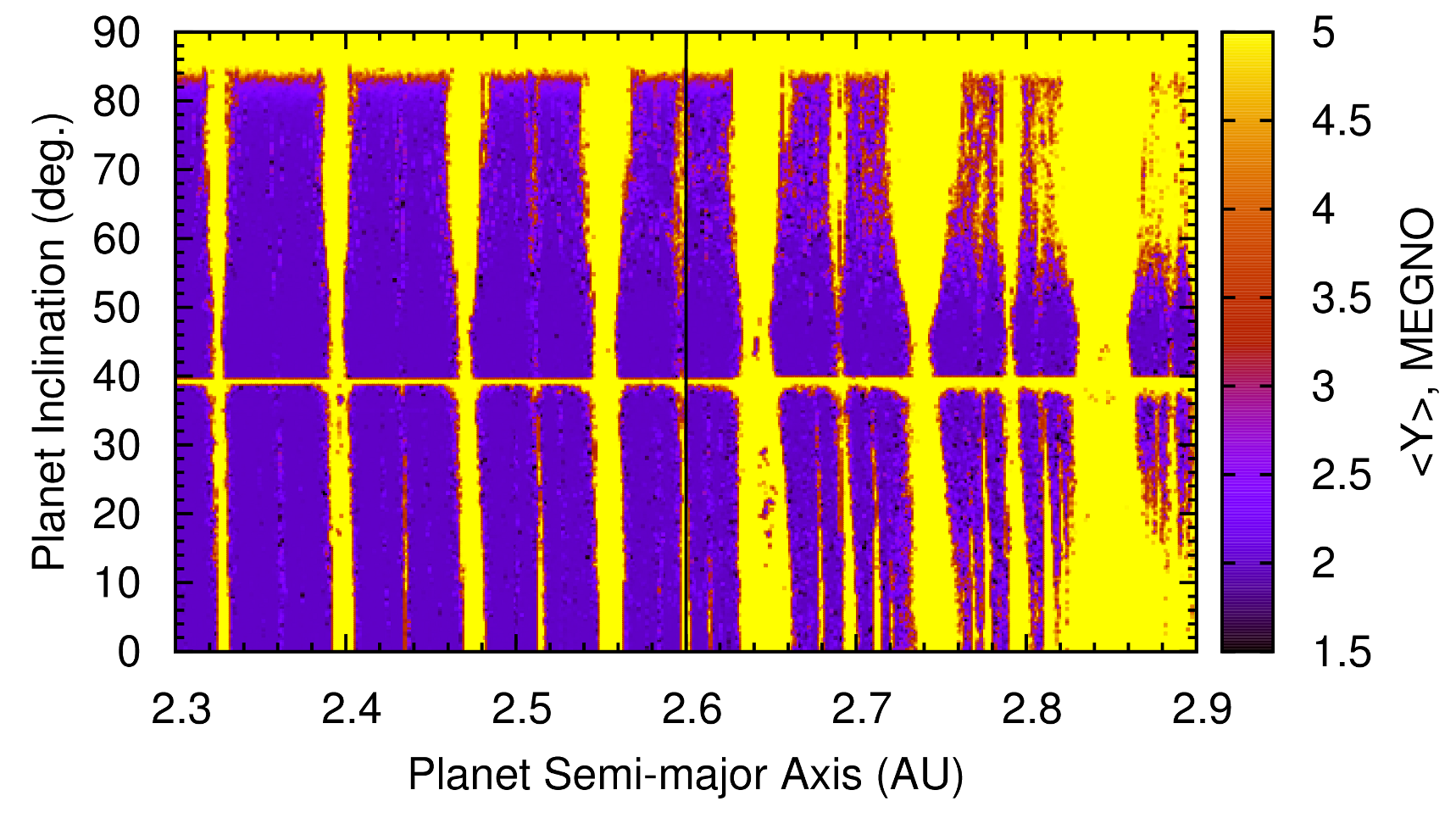}}
\subfloat{\includegraphics[width=.48\linewidth]{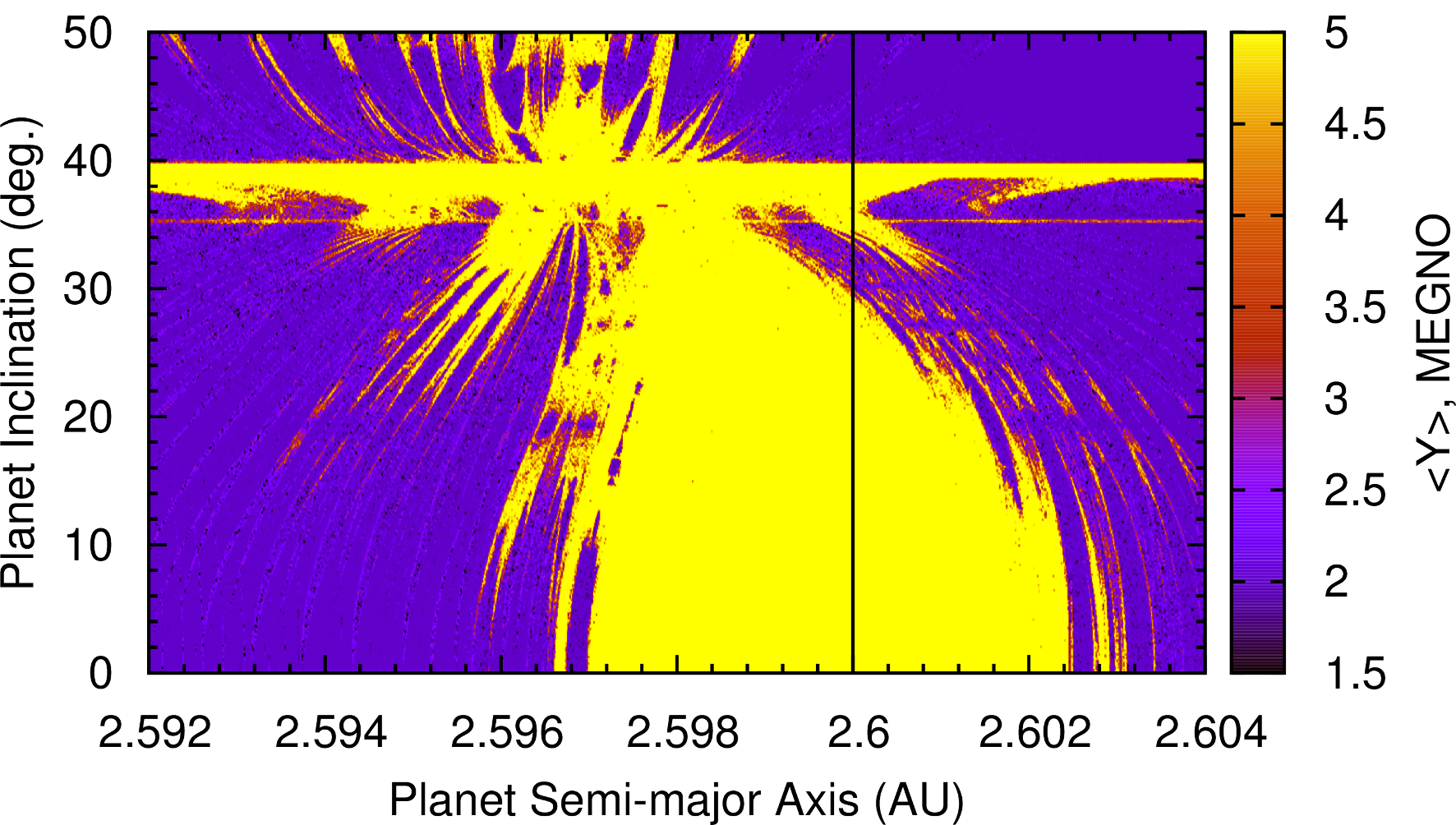}}
\caption{MEGNO maps indicating the quasi-periodic and chaotic regions of the giant planet in the HD 196885 system for inclination varying from (left) 0$^\circ$ to 90$^\circ$ and semimajor axis ranging from [2.3-2.9] au, simulated for 2 $\times 10^5$ years. Vertical line indicates the location of the best-fit semimajor axis of the planet at $a_{pl}$ = 2.6 au. A small area along the vertical line and for lower $i_{pl}$ region is magnified to clearly differentiate between the quasi-periodic and chaotic regions (right panel) and simulated for 8 $\times 10^5$ years. The planet's \emph{near} 39:2 MMR is located slightly left of the planet's best-fit location. The MMR stops when $i_{pl}$ reaches $\sim$ 30$^\circ$, with a small semi quasi-periodic region extending from 30$^\circ$ to 34$^\circ$ which we expect to vanish for longer simulation time. Yellow region signifies the chaotic orbits within the total simulation time and dark purple region signifies the periodic orbits. The colour bar indicates the strength in the value of MEGNO, $<y>$.}
\label{fig:MEGNO_isa}
\end{figure*}

\begin{figure}
\centering
\subfloat{\includegraphics[width=1\linewidth]{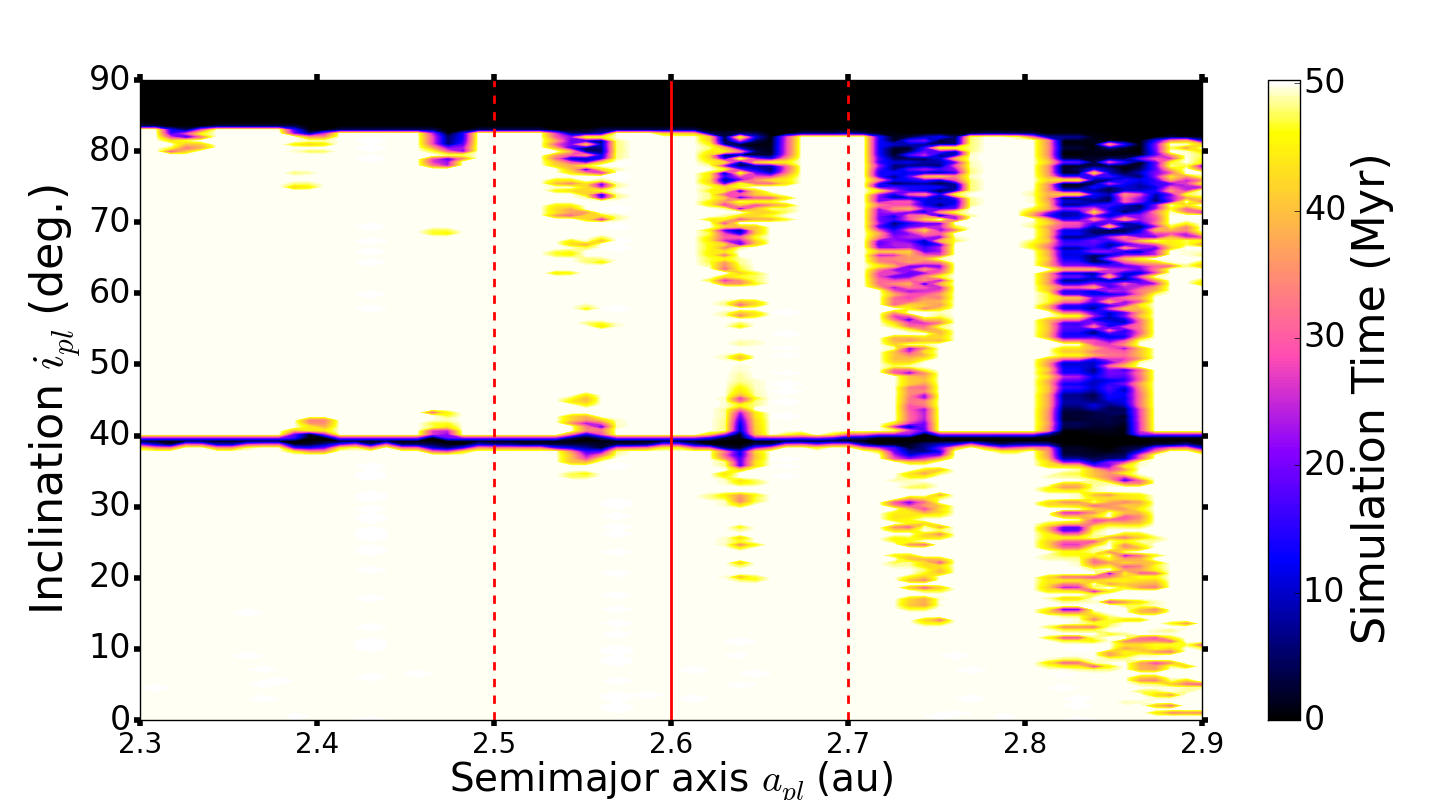}}
\caption{A global dynamical \emph{lifetime} map of the planet, HD 196885 Ab, for varying $i_{pl}$ and $a_{pl}$, simulated for 50 Myr. The colour bar indicates the survival time, where darker colour represents the instability (ejection or collision) and the lightest colour represent the stability (survival) up to the integration period. The solid red line at 2.6 au is the best-fit semimajor axis of the planet. Two dotted lines indicate the observational uncertainty of $\pm$0.1 au (see Table 1).  The vertical bars evident in Figs. \ref{fig:emax} and \ref{fig:MEGNO_isa} at 2.45, 2.55, 2.65, 2.75 au also appear.}
\label{fig:stability}
\end{figure}

We have generated the MEGNO maps considering a wide parameter space of the planet's initial inclination and semimajor axis which illustrates variations in the quasi-periodic and chaotic phase space for varying $i_{pl}$ and $a_{pl}$ values (Fig. \ref{fig:MEGNO_isa}). We found that, for $i_{pl}$ less than 39$^\circ$, the best-fit value of $a_{pl}$ (the vertical line at 2.6 au) lies within the chaotic region. However, it's important to note that a chaotic orbit does not necessarily imply an unstable one. This is shown above in Sec. \ref{sec:results} through the orbital integration of select initial conditions. Also, the global \emph{lifetime} map of $i_{pl}$ vs. $a_{pl}$ (Fig. \ref{fig:stability}) shows which initial conditions survive for the full integration time. The colour code in the \emph{lifetime} map is based on the planet's survival, ejection or collision time where the lightest colour represents the full survival of the planet. And the dark coloured vertical bands signifies instability, which means the planet at these initial conditions did not survive the total simulation time of 50 Myr.

The mean motion resonance (MMR) associated with the chaotic regions in the MEGNO maps can be estimated by calculating its position from the perturbation theory. For the best-fit values of the planet's and binary semimajor axes, and masses of the stars, the ratio of the ${(p+q)/p}$ commensurability can be calculated as (see \cite{mur99}),

\begin{align}
 \emph{p+q}\over{p} &= {\left({a_{bin}\over{a_{pl}}}\right)^{3\over{2}}{\left(m_{A}\over{m_{A} + m_{B}}\right)}^{1\over{2}}}, 
\label{eqn:resonance}
\end{align}

Where $m_{A}$ and $m_{B}$ represent masses of the primary and secondary star respectively; $q$ is the order of the resonance and $p$ is its degree. To search for the chaos associated with (p+q)/p MMR, we considered the resonant angle ($\Phi$) of the form

\begin{align}
 \emph{$\Phi$} &= {k_1\lambda_{bin} + k_2\lambda_{pl} + k_3\varpi_{bin} + k_4\varpi_{pl} + k_5\Omega_{bin} + k_6\Omega_{pl}},
\label{eqn:PHI}
\end{align}

Where the coefficients follow the relations {k$_1$} = {p+q}, {k$_2$} = -p,  and $\sum_{i}{k_i}$ = {0}.  The mean longitude $\lambda$ is a function of the mean anomaly, $M$, and longitude of pericenter, $\varpi$. Also, the sum of k$_5$ and k$_6$ is even as required by symmetry in the d'Alembert rules. 

\begin{figure*}
\centering
\subfloat[Planet's Inclination (i$_{pl}$) = 0$^\circ$]{\includegraphics[width=.35\linewidth]{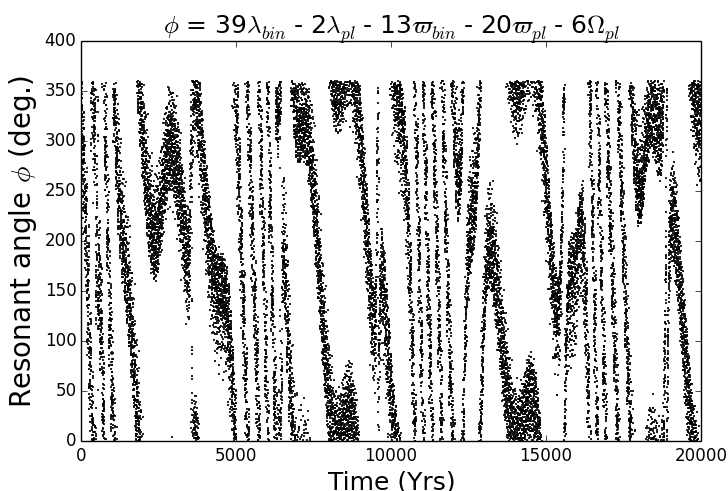}}
\subfloat[Planet's Inclination (i$_{pl}$) = 10$^\circ$]{\includegraphics[width=.35\linewidth]{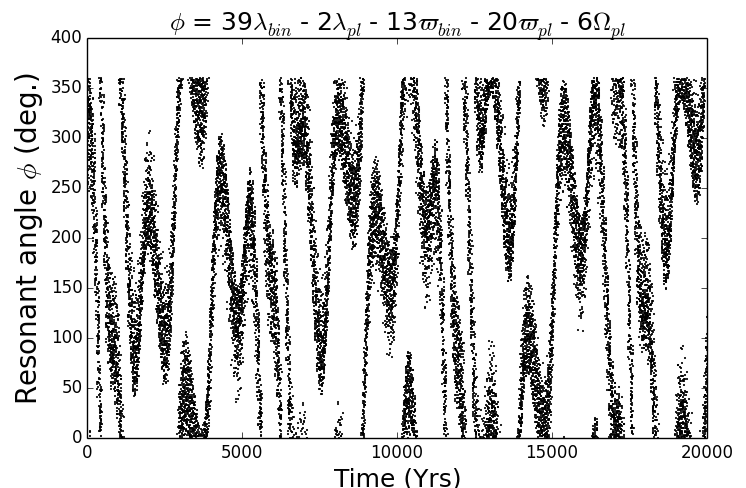}}
\subfloat[Planet's Inclination (i$_{pl}$) = 31$^\circ$]{\includegraphics[width=.35\linewidth]{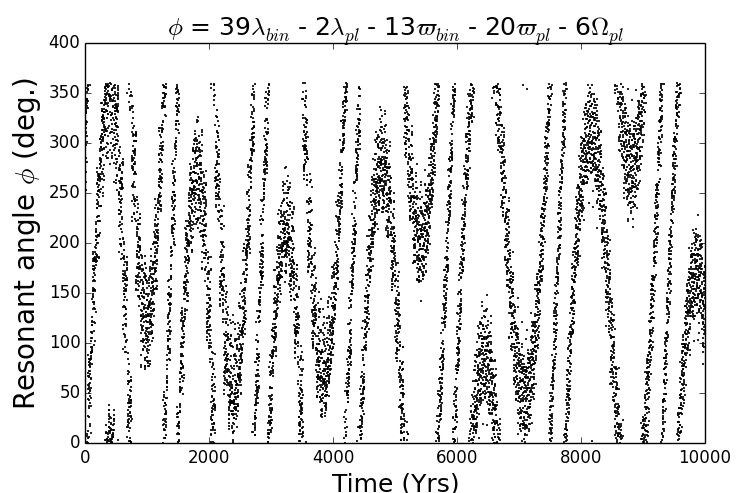}}
\\*
\subfloat[Planet's Inclination (i$_{pl}$) = 35$^\circ$]{\includegraphics[width=.35\linewidth]{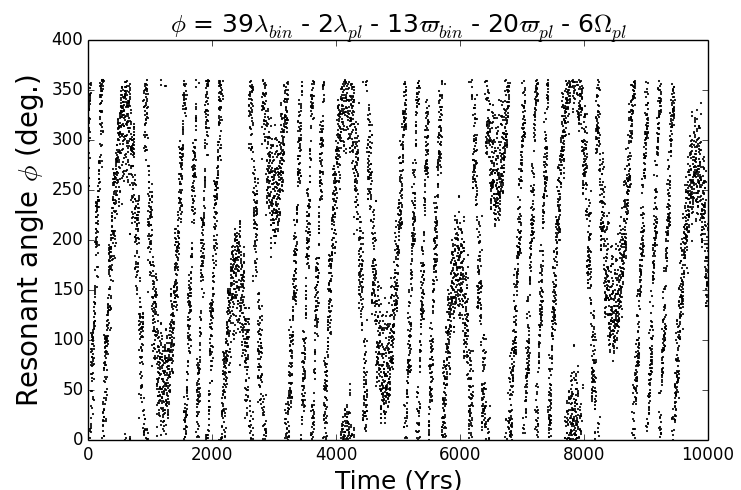}}
\subfloat[Planet's Inclination (i$_{pl}$) = 50$^\circ$]{\includegraphics[width=.35\linewidth]{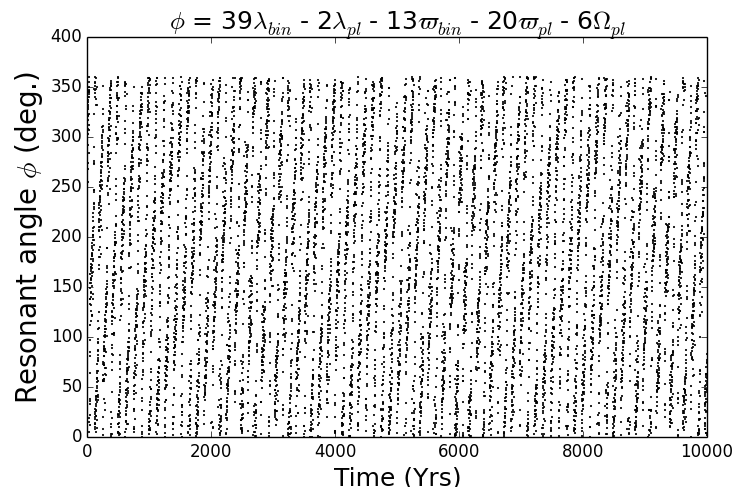}}
\subfloat[Planet's Inclination (i$_{pl}$) = 0$^\circ$,35$^\circ$]{\includegraphics[width= 6.25cm, height = 4.25cm]{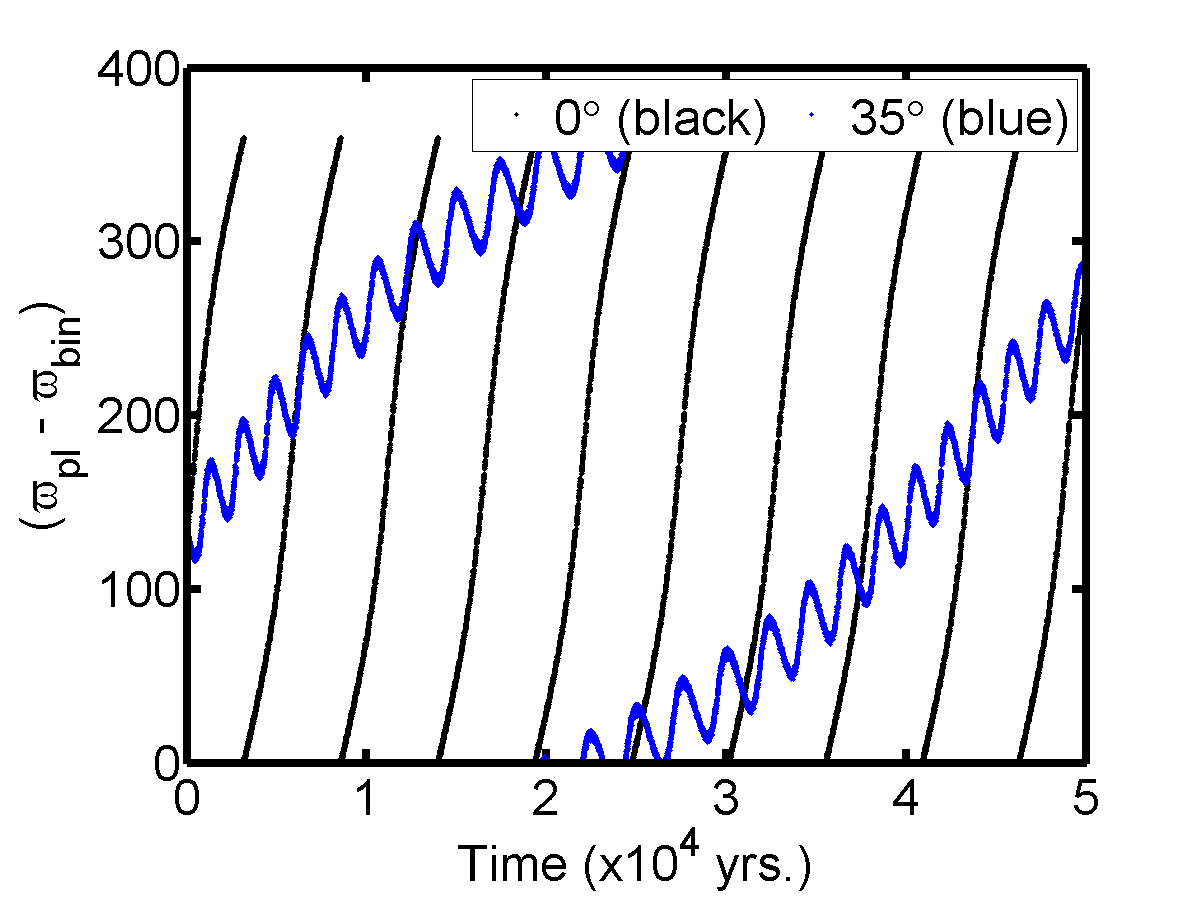}}
\caption{The time evolution of resonant angles ($\Phi$) calculated for different i$_{pl}$ values but along the best-fit location of semimajor axis ($a_{pl}$ = 2.6 au). The location for $\Phi$ angle test is chosen from MEGNO map (see Fig. \ref{fig:MEGNO_isa}) where measure of MEGNO ($<y>$) is quasi-periodic (e) and chaotic (a, b, c and d). (f) shows the time evolution of the angles ($\varpi_{pl} - \varpi_{bin}$) for $i_{pl}$ values at 0$^\circ$ and 35$^\circ$.}
\label{fig:PHI_1}
\end{figure*}

For the planet at 2.6 au from the primary component an interaction with the mean motion resonance (MMR) is found to be \emph{near} a 39:2 commensurability with the secondary component (some of the angles in Fig. \ref{fig:PHI_1} show a temporary lock). The time evolution of the angles $\varpi_{pl} - \varpi_{bin}$ (secular resonance, Fig.  \ref{fig:PHI_1},f) shows circulation between 0$^\circ$ and 360$^\circ$ when the $i_{pl}$ is set at lower values (for example 0$^\circ$) which suggests that the observed chaos is caused by the MMR. Secular resonance at the semi quasi-periodic region ($i_{pl}$ = 35$^\circ$) shows circulation as well, but with a  different orientation \citep{mur99}. Then, the MMR value is used to search for a librating resonant angle (Eq. \ref{eqn:PHI}), for k$_1$ = 39 , k$_2$ = -2, and various other combinations of k$_3$, k$_4$, k$_5$ and k$_6$ due to the influence of secular components. In our case, the $i_{pl}$ was varied with respect to the binary orbital plane and the orbits were integrated using astrocentric coordinates, considering the primary at the origin. Thus, k$_5$ in Eq. \ref{eqn:PHI} would be zero as $\Omega_{bin}$ becomes undefined ($i_{bin}$ = 0$^\circ$) and convention dictates that it be set to zero.

Initial conditions for different $i_{pl}$ values, but along the best-fit semimajor axis line, were picked where the measure of MEGNO is chaotic (Fig. \ref{fig:MEGNO_isa}, right panel) and the orbits were integrated for a short period with a high data sampling frequency. We would expect the resonance angle ($\Phi$) time series to alternate between modes of circulation and libration for initial conditions taken from the chaotic region and $\Phi$ to circulate only when the initial condition is taken from the quasi-periodic region. At 0$^\circ$, 10$^\circ$, 31$^\circ$ and 35$^\circ$, $\Phi$ is found to librate and circulate as expected for chaotic behaviour (Fig. \ref{fig:PHI_1}, a-d). This also confirms the earlier indication of the planet's interaction with a near 39:2 MMR with the secondary star. One of the factors inducing chaos in the region of low inclination orbit is possibly due to this near 39:2 MMR. The MMR in the case of higher inclination orbit changes its dynamical character which results in the periodic orbits. The location of the planet (vertical line at 2.6 au) is almost at the center of the MMR, thus minimizing the amplitude of the libration. The 39:2 MMR has a minimal effect for $i_{pl}$ greater than 40$^\circ$ and the planet enters a quasi-periodic region for $i_{pl}$ up to 55$^\circ$, a region where the $e_{max}$ deviates least from its best-fit value. The resonant angle is found to circulate only in these quasi-periodic regions. $\Phi$ plotted for initial conditions at 50$^\circ$ has shown circulation in Fig. \ref{fig:PHI_1}e. This gives a strong evidence that one of the factors responsible for the chaos below 39$^{\circ}$ is produced by the near 39:2 MMR interaction. Also, the secular perturbation induced by the precession of $\omega$ and $\Omega$ is found to contribute to the observed chaotic dynamics of the system in the Kozai regime. The libration and circulation of resonant angle seen in Fig. \ref{fig:PHI_1} is obtained only when the coefficients of $\varpi_{pl}$ and $\Omega_{pl}$ are included in the $\Phi$ term.

A small region of possible quasi-periodicity appears when the initial $i_{pl}$ lies between 30$^\circ$ and 34$^\circ$ (Fig. \ref{fig:MEGNO_isa} right panel). This would seem like a periodic region but the resonance angle test (Fig. \ref{fig:PHI_1}c) shows that the $\Phi$ is found to circulate and librate, which indicates this region as a chaotic zone. The small black and red dots in this region are indicative that many of these initial conditions have not had enough time to reveal their true nature (quasi-periodic or chaotic). What we called before a quasi-periodic region and expected the circulation of $\Phi$ has revealed itself as a region of complex dynamics with many overlapping and interacting resonances.

A strip of unstable chaos is observed when the initial $i_{pl}$ ranges between $\sim$(39$^\circ$ - 40$^\circ$) (Fig. \ref{fig:MEGNO_isa}, right panel) and for any choice of $a_{pl}$, from 2.3 au to 2.9 au. The instability at this region is also confirmed by the \emph{lifetime} map where the planet is found to collide with its host star. The MEGNO has been successful in determining the chaotic region surrounding this instability point. The minimum critical angle ($i_{pl} \ge 39.2^\circ$) required to initiate the Kozai mechanism lies within this inclination regime. Once the Kozai resonance phenomenon is triggered, it drives the planet into higher inclination orbits (hence higher eccentricity) causing the planet to collide with the primary. However, beyond 40$^\circ$ and less than 55$^\circ$ the planet maintains quasi-periodic orbits even though it is subject to the Kozai resonance. The observed horizontal stripe in the MEGNO map at $i_{pl}$ $\sim$ 39$^\circ$ can be considered as a separatrix line which separates the phase space from circulation when $i_{pl}$ $<$ 39.2$^\circ$, and libration when $i_{pl}$ $\ge$ 39.2$^\circ$.

The MEGNO maps have displayed quasi-periodic orbits along the best-fit $a_{pl}$ value ( black line, Fig. \ref{fig:MEGNO_isa}) and for $i_{pl}$ values greater than 40$^\circ$ and until it rises as high as $\sim$55$^\circ$ above the binary plane. Beyond this inclination regime, the maps have displayed the chaotic orbits. The MEGNO have shown an overall effectiveness at indicating the MMR locations, and the chaotic and periodic regions of the system's phase space. In addition, the dynamical maximum eccentricity and \emph{lifetime} maps have addressed the caveats of chaos seen in the MEGNO maps to discriminate which regions are truly unstable or merely chaotic within the integration period.

The planet's interaction near the 39:2 MMR with the secondary is observed when the mean best-fit value of semimajor axis is 2.6 au. When the planet's location is moved to 2.58 au and 2.68 au (but within the observational uncertainty limit), the planet is found to exhibit oscillations due to the 20:1 and 19:1 MMRs, respectively.

\begin{figure}
\centering
\hspace*{\fill}
\subfloat{\includegraphics[width=1\linewidth]{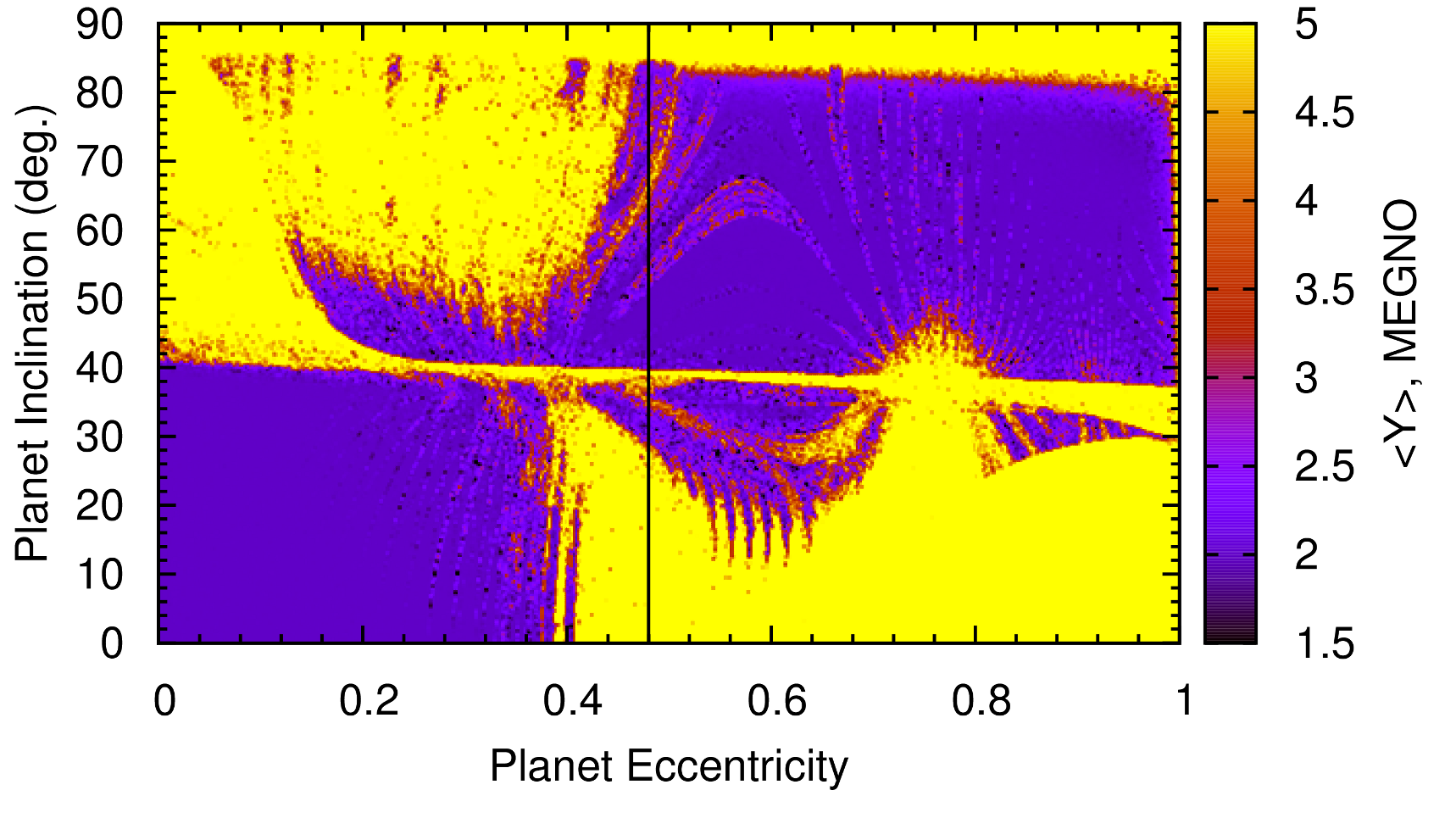}}\hfill
\hspace*{\fill}%
\caption{MEGNO maps indicating the periodic and chaotic orbits of the giant planet in HD 196885 for inclination varying from 0$^\circ$ to 90$^\circ$, plotted versus the eccentricity and simulated for 2 $\times 10^5$ years. Vertical line indicates the planet's best-fit location at $e_{pl}$ = 0.48.}
\label{fig:MEGNO_ie}
\end{figure}

The planet's orbital motion is also analysed in the varying $i_{pl}$ and $e_{pl}$ plane (Fig. \ref{fig:MEGNO_ie}) to display possible periodic and chaotic regimes. The vertical line at $e_{pl}$ = 0.48 indicates the planet's best-fit location, and, like in the previous case (Fig. \ref{fig:MEGNO_isa}), the planet clearly lies in chaotic zone for $e_{pl}$ $>$ 0.4 and $i_{pl}$ $<$ 30$^\circ$. Driven by the Kozai resonance, a similar chaotic stripe appears at $i_{pl}$ $\sim$ 40$^\circ$. Periodic regions continue beyond this angle, however some chaotic islands do appear for $i_{pl}$ greater than 55$^\circ$.

The discovery of the planet in the HD 196885 system \citep{cha11} using the radial velocity (RV) technique has only allowed the determination of the planet's minimum mass of 2.98$M_J$. We have performed a test to indicate the periodic and/or chaotic regions of the system for variations in the planet's mass ($m_{pl}$) ranging from 2$M_J$ to 20$M_J$ and  $i_{pl}$ varying from 0$^\circ$ to 90$^\circ$ (Fig. \ref{fig:MEGNO_im}). The map indicates that quasi-periodic regions exist for various combination of $m_{pl}$ and $i_{pl}$. The system displays periodic orbits for $m_{pl}$ mostly between 3$M_J$ and 6$M_J$ with a small chaotic region around 4.5$M_J$ plus the exclusion of chaotic region at $\sim$ 40$^\circ$. The chaotic region continues below 30$^\circ$ at the best-fit semimajor axis location of the planet. The map indicates other possible periodic islands for higher mass values of the planet specifically for $i_{pl}$ between (40$^\circ$ - 50$^\circ$) and $m_{pl}$ between $\sim$ (8.5 - 11.5)$M_J$, (12.5 - 15)$M_J$ and (16 - 19)$M_J$.


\section{Conclusions}

Based on our analysis, the best configuration for the planet to display quasi-periodic orbits is when the planet's orbital inclination relative to the binary plane lies between 41$^\circ$ and 55$^\circ$. \cite{cha11} made a similar suggestion regarding the planet's inclination that a high relative inclination favours for the periodic orbits. Also, \cite{giu12} have found that the planet has either coplanar orbits (prograde and retrograde) or highly inclined orbits near the \emph{Lido-Kozai} equilibrium points, $i$ - 90$^\circ$ = $\pm$47$^\circ$. This range of angles gets smaller when the planet's mass is increased more than the current value of 2.98$M_{J}$. The planet can have an interaction with the near 39:2 MMR from perturbations of the secondary component of the binary which appears to be responsible for the chaotic region as seen in the MEGNO maps below 39$^\circ$. The plots of the resonance angle time series for the near 39:2 MMR clearly demonstrate the circulation and libration of $\Phi$ based on the choices of initial condition. The planet's orbital configuration below 39$^\circ$ is proven chaotic, and hence lesser possibility for the planet to be in that regime under the current assumptions of planetary formation.

The best-fit value of the planet's eccentricity was set at 0.48. Then the system was evolved for 50Myr and the maximum $e_{pl}$ value was calculated in $a_{pl}$ vs. $i_{pl}$ parameter space. The $e_{pl}$ is found to deviate least from its best-fit value when $i_{pl}$ is at $\sim$35$^\circ$ and $\sim$(40$^\circ$-50$^\circ$), thus, these regions are expected to demonstrate the most quasi-periodic regimes for the planet. The high amplitude oscillation in $e_{pl}$ below 35$^\circ$ mainly arises due to the planet's secular interactions with the secondary. The perturbation from the secondary may have forced the planetary embryo or protoplanet into precession during its early formation stage, eventually driving the planet into high inclination orbits.

\begin{figure}
\centering
\hspace*{\fill}
\subfloat{\includegraphics[width=1\linewidth]{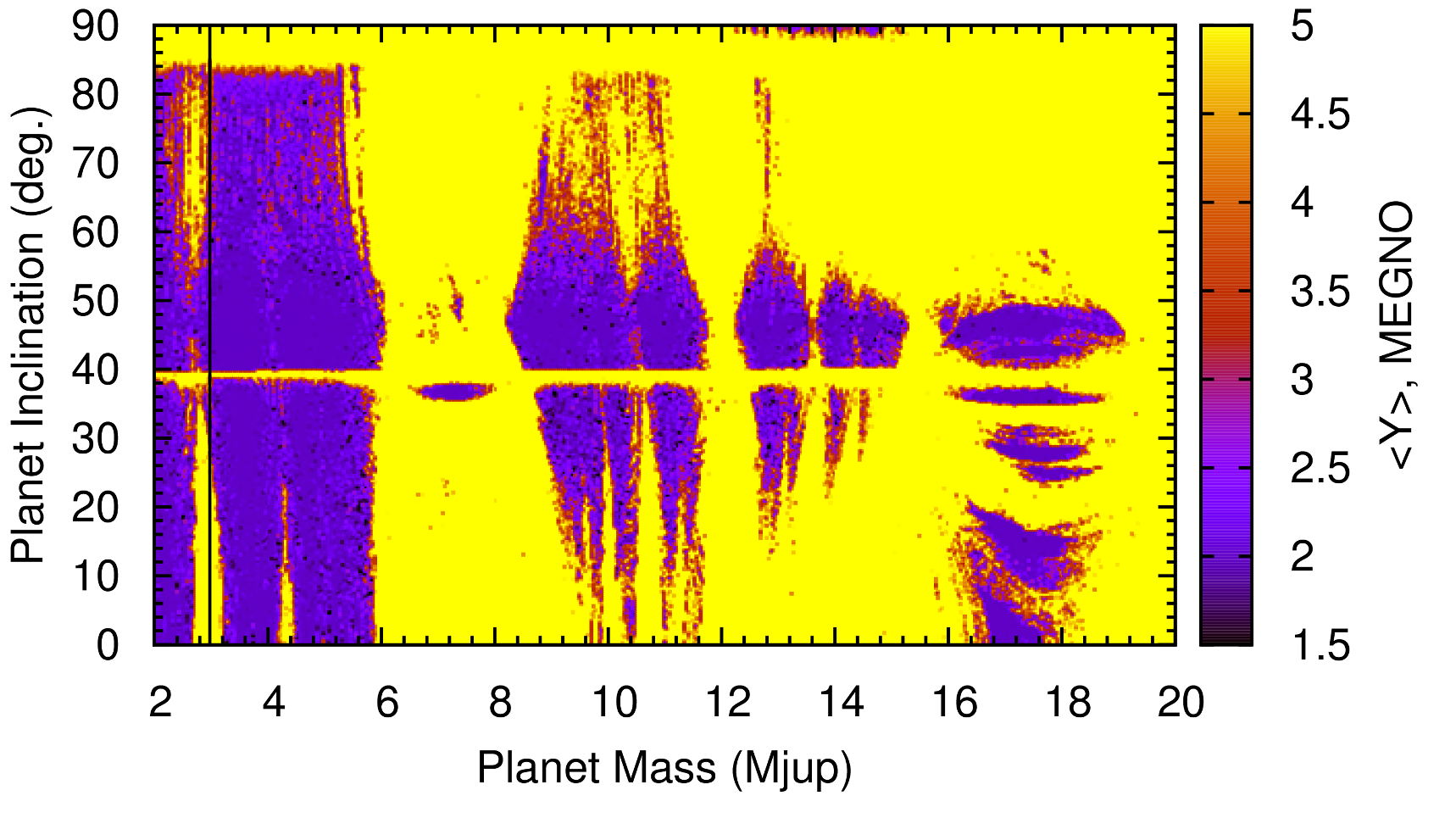}}
\hspace*{\fill}%
\caption{MEGNO maps indicating the periodic and chaotic orbits of the giant planet in HD 196885 for inclination varying from 0$^\circ$ to 90$^\circ$, plotted versus the planet's mass and simulated for 2 $\times 10^5$ years. Vertical line indicates the planet's best-fit location at $m_{pl}$ = 2.98$M_J$.}
\label{fig:MEGNO_im}
\end{figure}

The planet’s higher mass is possibly constrained between (3 - 6)$M_{J}$; however, these choices also depend on the choice of planet's orbital inclination and it is mostly favorable when $m_{pl}$ is less than 6$M_{J}$ and $i_{pl}$ less than $\sim 55^\circ$, with some exceptions of chaotic islands (Fig. \ref{fig:MEGNO_im}). The planetary mass higher than 9$M_{J}$ and up to 19$M_{J}$ is likely when the planet's orbital inclination lies in a smaller regime, somewhere between 40$^\circ$ and 50$^\circ$. A possibility of having any other terrestrial planets within the 2.6 au region around the primary is less likely. The circumprimary disc can be strongly hostile to planetesimal accretion in this region \citep{the11}. \\

\emph{\textbf{Acknowledgement}}: SS and JN would like to thank The Department of Physics at UT Arlington, Zdzislaw Musielak and Manfred Cuntz for their continuous support and guidance. BQ gratefully acknowledges support from the NASA post doctoral program. TCH gratefully acknowledges financial support from the Korea Research Council for Fundamental Science and Technology (KRCF) through the Young Research Scientist Fellowship Program and financial support from KASI (Korea Astronomy and Space Science Institute) grant number 2014-1-400-06. Numerical computations were partly carried out using the SFI/HEA Irish Centre for High-End Computing (ICHEC) and the PLUTO computing cluster at the Korea Astronomy and Space Science Institute. Astronomical research at the Armagh Observatory is funded by the Northern Ireland Department of Culture, Arts and Leisure (DCAL). We would also like to thank the anonymous referee for his/her comments and suggestions on eccentric \emph{Lidov-Kozai} mechanism which significantly improved this work.

\bibliographystyle{mn2e}
\bibliography{references}

\begin{thebibliography}{}

\bibitem[\protect\citeauthoryear{{Antonini} \& {Merritt}}{{Antonini} \&
  {Merritt}}{2013}]{ant13}
{Antonini} F.,  {Merritt} D.,  2013, \apjl, 763, L10

\bibitem[\protect\citeauthoryear{{Chambers}}{{Chambers}}{1999}]{cha99}
{Chambers} J.~E.,  1999, \mnras, 304, 793

\bibitem[\protect\citeauthoryear{{Chambers} \& {Migliorini}}{{Chambers} \&
  {Migliorini}}{1997}]{cha97}
{Chambers} J.~E.,  {Migliorini} F.,  1997, in AAS/Division for Planetary
  Sciences Meeting Abstracts \#29 Vol.~29 of Bulletin of the American
  Astronomical Society, {Mercury - A New Software Package for Orbital
  Integrations}.
p.~1024

\bibitem[\protect\citeauthoryear{{Chatterjee}, {Ford}, {Matsumura} \&
  {Rasio}}{{Chatterjee} et~al.}{2008}]{cha08}
{Chatterjee} S.,  {Ford} E.~B.,  {Matsumura} S.,    {Rasio} F.~A.,  2008, \apj,
  686, 580

\bibitem[\protect\citeauthoryear{{Chauvin}, {Beust}, {Lagrange} \&
  {Eggenberger}}{{Chauvin} et~al.}{2011}]{cha11}
{Chauvin} G.,  {Beust} H.,  {Lagrange} A.-M.,    {Eggenberger} A.,  2011, \aap,
  528, A8

\bibitem[\protect\citeauthoryear{{Cincotta} \& {Sim{\'o}}}{{Cincotta} \&
  {Sim{\'o}}}{1999}]{cin99}
{Cincotta} P.,  {Sim{\'o}} C.,  1999, Celestial Mechanics and Dynamical
  Astronomy, 73, 195

\bibitem[\protect\citeauthoryear{{Cincotta} \& {Sim{\'o}}}{{Cincotta} \&
  {Sim{\'o}}}{2000}]{cin00}
{Cincotta} P.~M.,  {Sim{\'o}} C.,  2000, aaps, 147, 205

\bibitem[\protect\citeauthoryear{{Correia}, {Laskar}, {Farago} \&
  {Bou{\'e}}}{{Correia} et~al.}{2011}]{cor11}
{Correia} A.~C.~M.,  {Laskar} J.,  {Farago} F.,    {Bou{\'e}} G.,  2011,
  Celestial Mechanics and Dynamical Astronomy, 111, 105

\bibitem[\protect\citeauthoryear{{Cuntz}, {Quarles}, {Eberle} \&
  {Shukayr}}{{Cuntz} et~al.}{2013}]{cun13}
{Cuntz} M.,  {Quarles} B.,  {Eberle} J.,    {Shukayr} A.,  2013, \pasa, 30, 33

\bibitem[\protect\citeauthoryear{{Fabrycky} \& {Tremaine}}{{Fabrycky} \&
  {Tremaine}}{2007}]{fab07}
{Fabrycky} D.,  {Tremaine} S.,  2007, \apj, 669, 1298

\bibitem[\protect\citeauthoryear{{Ford}, {Kozinsky} \& {Rasio}}{{Ford}
  et~al.}{2000}]{for00}
{Ford} E.~B.,  {Kozinsky} B.,    {Rasio} F.~A.,  2000, \apj, 535, 385

\bibitem[\protect\citeauthoryear{{Giuppone}, {Morais}, {Bou{\'e}} \&
  {Correia}}{{Giuppone} et~al.}{2012}]{giu12}
{Giuppone} C.~A.,  {Morais} M.~H.~M.,  {Bou{\'e}} G.,    {Correia} A.~C.~M.,
  2012, \aap, 541, A151

\bibitem[\protect\citeauthoryear{{Go{\'z}dziewski}, {Bois}, {Maciejewski} \&
  {Kiseleva-Eggleton}}{{Go{\'z}dziewski} et~al.}{2001}]{goz01b}
{Go{\'z}dziewski} K.,  {Bois} E.,  {Maciejewski} A.~J.,    {Kiseleva-Eggleton}
  L.,  2001, \aap, 378, 569

\bibitem[\protect\citeauthoryear{{Go{\'z}dziewski} \&
  {Maciejewski}}{{Go{\'z}dziewski} \& {Maciejewski}}{2001}]{goz01a}
{Go{\'z}dziewski} K.,  {Maciejewski} A.~J.,  2001, \apjl, 563, L81

\bibitem[\protect\citeauthoryear{{Heller}}{{Heller}}{2012}]{hel12}
{Heller} R.,  2012, \aap, 545, L8

\bibitem[\protect\citeauthoryear{{Hinse}, {Christou}, {Alvarellos} \&
  {Go{\'z}dziewski}}{{Hinse} et~al.}{2010}]{hin10}
{Hinse} T.~C.,  {Christou} A.~A.,  {Alvarellos} J.~L.~A.,    {Go{\'z}dziewski}
  K.,  2010, \mnras, 404, 837

\bibitem[\protect\citeauthoryear{{Hinse}, {Michelsen}, {J{\o}rgensen},
  {Go{\'z}dziewski} \& {Mikkola}}{{Hinse} et~al.}{2008}]{hin08}
{Hinse} T.~C.,  {Michelsen} R.,  {J{\o}rgensen} U.~G.,  {Go{\'z}dziewski} K.,
   {Mikkola} S.,  2008, \aap, 488, 1133

\bibitem[\protect\citeauthoryear{{Holman}, {Touma} \& {Tremaine}}{{Holman}
  et~al.}{1997}]{hol97}
{Holman} M.,  {Touma} J.,    {Tremaine} S.,  1997, \nat, 386, 254

\bibitem[\protect\citeauthoryear{{Innanen}, {Zheng}, {Mikkola} \&
  {Valtonen}}{{Innanen} et~al.}{1997}]{inn97}
{Innanen} K.~A.,  {Zheng} J.~Q.,  {Mikkola} S.,    {Valtonen} M.~J.,  1997,
  \aj, 113, 1915

\bibitem[\protect\citeauthoryear{{Katz}, {Dong} \& {Malhotra}}{{Katz}
  et~al.}{2011}]{kat11}
{Katz} B.,  {Dong} S.,    {Malhotra} R.,  2011, Physical Review Letters, 107,
  181101

\bibitem[\protect\citeauthoryear{{Kipping}, {Hartman}, {Buchhave}, {Schmitt},
  {Bakos} \& {Nesvorn{\'y}}}{{Kipping} et~al.}{2013}]{kip13}
{Kipping} D.~M.,  {Hartman} J.,  {Buchhave} L.~A.,  {Schmitt} A.~R.,  {Bakos}
  G.~{\'A}.,    {Nesvorn{\'y}} D.,  2013, \apj, 770, 101

\bibitem[\protect\citeauthoryear{{Kozai}}{{Kozai}}{1962}]{koz62}
{Kozai} Y.,  1962, \aj, 67, 591

\bibitem[\protect\citeauthoryear{{Li}, {Naoz}, {Kocsis} \& {Loeb}}{{Li}
  et~al.}{2013}]{li14}
{Li} G.,  {Naoz} S.,  {Kocsis} B.,    {Loeb} A.,  2013, ArXiv e-prints

\bibitem[\protect\citeauthoryear{{Lidov}}{{Lidov}}{1962}]{lid62}
{Lidov} M.~L.,  1962, \planss, 9, 719

\bibitem[\protect\citeauthoryear{{Lithwick} \& {Naoz}}{{Lithwick} \&
  {Naoz}}{2011}]{lit11}
{Lithwick} Y.,  {Naoz} S.,  2011, \apj, 742, 94

\bibitem[\protect\citeauthoryear{Lyapunov}{Lyapunov}{1907}]{lya07}
Lyapunov A.,  1907, Annales de la facult\`e des sciences de Toulouse, 2:9, 203

\bibitem[\protect\citeauthoryear{{Mayor} \& {Queloz}}{{Mayor} \&
  {Queloz}}{1995}]{may95}
{Mayor} M.,  {Queloz} D.,  1995, \nat, 378, 355

\bibitem[\protect\citeauthoryear{{Mikkola} \& {Innanen}}{{Mikkola} \&
  {Innanen}}{1999}]{mik99}
{Mikkola} S.,  {Innanen} K.,  1999, Celestial Mechanics and Dynamical
  Astronomy, 74, 59

\bibitem[\protect\citeauthoryear{{Murray} \& {Dermott}}{{Murray} \&
  {Dermott}}{1999}]{mur99}
{Murray} C.~D.,  {Dermott} S.~F.,  1999, {Solar system dynamics}

\bibitem[\protect\citeauthoryear{{Nagasawa}, {Ida} \& {Bessho}}{{Nagasawa}
  et~al.}{2008}]{nag08}
{Nagasawa} M.,  {Ida} S.,    {Bessho} T.,  2008, \apj, 678, 498

\bibitem[\protect\citeauthoryear{{Naoz}, {Farr}, {Lithwick}, {Rasio} \&
  {Teyssandier}}{{Naoz} et~al.}{2011}]{nao11}
{Naoz} S.,  {Farr} W.~M.,  {Lithwick} Y.,  {Rasio} F.~A.,    {Teyssandier} J.,
  2011, \nat, 473, 187

\bibitem[\protect\citeauthoryear{{Naoz}, {Farr}, {Lithwick}, {Rasio} \&
  {Teyssandier}}{{Naoz} et~al.}{2013}]{nao13}
{Naoz} S.,  {Farr} W.~M.,  {Lithwick} Y.,  {Rasio} F.~A.,    {Teyssandier} J.,
  2013, \mnras, 431, 2155

\bibitem[\protect\citeauthoryear{{Naoz}, {Farr} \& {Rasio}}{{Naoz}
  et~al.}{2012}]{nao12}
{Naoz} S.,  {Farr} W.~M.,    {Rasio} F.~A.,  2012, \apjl, 754, L36

\bibitem[\protect\citeauthoryear{{Naoz}, {Kocsis}, {Loeb} \& {Yunes}}{{Naoz}
  et~al.}{2013}]{nao13a}
{Naoz} S.,  {Kocsis} B.,  {Loeb} A.,    {Yunes} N.,  2013, \apj, 773, 187

\bibitem[\protect\citeauthoryear{{Noyola}, {Satyal} \& {Musielak}}{{Noyola}
  et~al.}{2013}]{noy13}
{Noyola} J.~P.,  {Satyal} S.,    {Musielak} Z.~E.,  2013, ArXiv e-prints

\bibitem[\protect\citeauthoryear{{Quarles}, {Eberle}, {Musielak} \&
  {Cuntz}}{{Quarles} et~al.}{2011}]{qua11}
{Quarles} B.,  {Eberle} J.,  {Musielak} Z.~E.,    {Cuntz} M.,  2011, A\&A, 533,
  A2

\bibitem[\protect\citeauthoryear{{Rasio} \& {Ford}}{{Rasio} \&
  {Ford}}{1996}]{ras96}
{Rasio} F.~A.,  {Ford} E.~B.,  1996, Science, 274, 954

\bibitem[\protect\citeauthoryear{{Ricker}, {Latham}, {Vanderspek}, {Ennico},
  {Bakos}, {Brown}, {Burgasser}, {Charbonneau}, {Clampin}, {Deming}, {Doty},
  {Dunham}, {Elliot}, {Holman}, {Ida} \& {Jenkins}}{{Ricker}
  et~al.}{2010}]{ric10}
{Ricker} G.~R.,  {Latham} D.~W.,  {Vanderspek} R.~K.,  {Ennico} K.~A.,  {Bakos}
  G.,  {Brown} T.~M.,  {Burgasser} A.~J.,  {Charbonneau} D.,  {Clampin} M.,
  {Deming} L.~D.,  {Doty} J.~P.,  {Dunham} E.~W.,  {Elliot} J.~L.,  {Holman}
  M.~J.,  {Ida} S.,    {Jenkins} J.~M.,  2010, in American Astronomical Society
  Meeting Abstracts \#215 Vol.~42 of Bulletin of the American Astronomical
  Society, {Transiting Exoplanet Survey Satellite (TESS)}.
p. 450.06

\bibitem[\protect\citeauthoryear{{Satyal}, {Quarles} \& {Hinse}}{{Satyal}
  et~al.}{2013}]{sat13}
{Satyal} S.,  {Quarles} B.,    {Hinse} T.~C.,  2013, \mnras, 433, 2215

\bibitem[\protect\citeauthoryear{{S{\l}onina}, {Go{\'z}dziewski} \&
  {Migaszewski}}{{S{\l}onina} et~al.}{2012}]{slo12}
{S{\l}onina} M.,  {Go{\'z}dziewski} K.,    {Migaszewski} C.,  2012, in {Arenou}
  F.,  {Hestroffer} D.,  eds, Proceedings of the workshop ''Orbital Couples:
  Pas de Deux in the Solar System and the Milky Way''. Held at the Observatoire
  de Paris, 10-12 October 2011. Editors: F. Arenou, D. Hestroffer. ISBN
  2-910015-64-5, p. 125-129 {Mechanic: a new numerical MPI framework for the
  dynamical astronomy}.
pp 125--129

\bibitem[\protect\citeauthoryear{{S{\l}onina}, {Go{\'z}dziewski} \&
  {Migaszewski}}{{S{\l}onina} et~al.}{2014}]{slo14}
{S{\l}onina} M.,  {Go{\'z}dziewski} K.,    {Migaszewski} C.,  2014, ArXiv
  e-prints

\bibitem[\protect\citeauthoryear{{Szenkovits} \& {Mak{\'o}}}{{Szenkovits} \&
  {Mak{\'o}}}{2008}]{sze08}
{Szenkovits} F.,  {Mak{\'o}} Z.,  2008, Celestial Mechanics and Dynamical
  Astronomy, 101, 273

\bibitem[\protect\citeauthoryear{{Teyssandier}, {Naoz}, {Lizarraga} \&
  {Rasio}}{{Teyssandier} et~al.}{2013}]{tey13}
{Teyssandier} J.,  {Naoz} S.,  {Lizarraga} I.,    {Rasio} F.~A.,  2013, \apj,
  779, 166

\bibitem[\protect\citeauthoryear{{Thebault}}{{Thebault}}{2011}]{the11}
{Thebault} P.,  2011, Celestial Mechanics and Dynamical Astronomy, 111, 29

\bibitem[\protect\citeauthoryear{{Veras} \& {Ford}}{{Veras} \&
  {Ford}}{2010}]{ver10}
{Veras} D.,  {Ford} E.~B.,  2010, \apj, 715, 803

\bibitem[\protect\citeauthoryear{{Winn}, {Fabrycky}, {Albrecht} \&
  {Johnson}}{{Winn} et~al.}{2010}]{win10}
{Winn} J.~N.,  {Fabrycky} D.,  {Albrecht} S.,    {Johnson} J.~A.,  2010, \apjl,
  718, L145

\bibitem[\protect\citeauthoryear{{Wu} \& {Murray}}{{Wu} \&
  {Murray}}{2003}]{wu03}
{Wu} Y.,  {Murray} N.,  2003, \apj, 589, 605

\bibitem[\protect\citeauthoryear{{Wu}, {Murray} \& {Ramsahai}}{{Wu}
  et~al.}{2007}]{wu07}
{Wu} Y.,  {Murray} N.~W.,    {Ramsahai} J.~M.,  2007, \apj, 670, 820

\end{thebibliography}

\end{document}